\documentclass[a4paper,journal]{IEEEtran}

\usepackage{amsmath,amsfonts,amssymb}
\usepackage{algpseudocode}
\usepackage{algorithm,mathtools}
\usepackage{array,multirow}
\usepackage{textcomp,enumitem}
\usepackage{stfloats}
\usepackage{ifthen}
\usepackage{url}
\usepackage{verbatim}
\usepackage{graphicx,caption,subcaption}
\usepackage[dvipsnames]{xcolor}
\usepackage{cite}
\usepackage{pgfplots,soul}
\usetikzlibrary{calc,fit,patterns,decorations.markings,matrix,3d}
\usepgfplotslibrary{colormaps,fillbetween}
\sethlcolor{SpringGreen}

\makeatletter
\def\customrevertcolormap#1{
	\pgfplotsarraycopy{pgfpl@cm@#1}\to{custom@COPY}
	\c@pgf@counta=0
	\c@pgf@countb=\pgfplotsarraysizeof{custom@COPY}\relax
	\c@pgf@countd=\c@pgf@countb
	\advance\c@pgf@countd by-1 
	\pgfutil@loop
	\ifnum\c@pgf@counta<\c@pgf@countb
	\pgfplotsarrayselect{\c@pgf@counta}\of{custom@COPY}\to\pgfplots@loc@TMPa
	\pgfplotsarrayletentry\c@pgf@countd\of{pgfpl@cm@#1}=\pgfplots@loc@TMPa
	\advance\c@pgf@counta by1 
	\advance\c@pgf@countd by-1 
	\pgfutil@repeat
}

\pgfplotsset{compat=newest,tick label style={font=\small},
	tick scale binop=\times,
	colormap={slategraywhite}{
		rgb255=(255,255,255)
		rgb255=(255,0,0)},
	every axis/.append style={label style={font=\small}}}

\newcommand{\ignore}[1]{}

\newcommand{\Fxf}{F_{x\rm{f}}}
\newcommand{\Fxr}{F_{x\rm{r}}}
\newcommand{\Fyf}{F_{y\rm{f}}}
\newcommand{\Fyr}{F_{y\rm{r}}}
\newcommand{\Fzf}{F_{z\rm{f}}}
\newcommand{\Fzr}{F_{z\rm{r}}}
\newcommand{\alf}{\alpha_{\rm{f}}}
\newcommand{\alr}{\alpha_{\rm{r}}}
\newcommand{\als}{\alpha_{\rm{s}}}
\newcommand{\lf}{l_{\rm{f}}}
\newcommand{\lr}{l_{\rm{r}}}
\newcommand{\e}{_{\rm{ego}}}
\newcommand{\obs}{_{\rm{obs}}}
\newcommand{\rf}{_{\mathrm{ref}}}
\newcommand{\No}{N_{\mathrm{o}}}
\newcommand{\Np}{N_{\mathrm{p}}}
\newcommand{\mmps}{_{\rm{MMPS}}}

\newtheorem{rem}{Remark}

\begin{document}

\title{Proactive Emergency Collision Avoidance for Automated Driving in Highway Scenarios}

\author{Leila~Gharavi,
	Azita~Dabiri,
	Jelske~Verkuijlen,
	Bart~De~Schutter,~\IEEEmembership{Fellow,~IEEE},
	and~Simone~Baldi,~\IEEEmembership{Senior~Member,~IEEE}
	\thanks{Leila~Gharavi, Azita~Dabiri, Jelske~Verkuijlen and Bart~De~Schutter are with the Delft Center for Systems and Control, Delft University of Technology, 2628 CD Delft, The Netherlands (e-mails: \texttt{L.Gharavi@tudelft.nl}; \texttt{A.Dabiri@tudelft.nl}; \texttt{J.R.M.Verkuijlen@student.tudelft.nl}; \texttt{B.DeSchutter@tudelft.nl}).}
	\thanks{Simone~Baldi is with the School of Mathematics, Southeast University, Nanjing 21118, China (e-mail: \texttt{SimoneBaldi@seu.edu.cn}).}}

\markboth{IEEE Transactions on Control Systems Technology,~Special~Issue,~September~2024}%
{Shell \MakeLowercase{\textit{et al.}}: Bare Demo of IEEEtran.cls for IEEE Journals}

\maketitle


\begin{abstract}
Uncertainty in the behavior of other traffic participants is a crucial factor in collision avoidance for automated driving; here, stochastic metrics could avoid overly conservative decisions. This paper introduces a Stochastic Model Predictive Control (SMPC) planner for emergency collision avoidance in highway scenarios to proactively minimize collision risk while ensuring safety through chance constraints. To guarantee that the emergency trajectory can be attained, we incorporate nonlinear tire dynamics in the prediction model of the ego vehicle. Further, we exploit Max-Min-Plus-Scaling (MMPS) approximations of the nonlinearities to avoid conservatism, enforce proactive collision avoidance, and improve computational efficiency in terms of performance and speed. Consequently, our contributions include integrating a dynamic ego vehicle model into the SMPC planner, introducing the MMPS approximation for real-time implementation in emergency scenarios, and integrating SMPC with hybridized chance constraints and risk minimization. We evaluate our SMPC formulation in terms of proactivity and efficiency in various hazardous scenarios. Moreover, we demonstrate the effectiveness of our proposed approach by comparing it with a state-of-the-art SMPC planner and we validate that the generated
trajectories can be attained using a high-fidelity vehicle model in IPG CarMaker. 
\end{abstract}

\begin{IEEEkeywords}
	Stochastic model predictive control, Emergency collision avoidance, Hybrid approximation, Highway automated driving
\end{IEEEkeywords}


\section{Introduction}\label{sec:intro}

\begin{figure}[hbt]\centering
	\includegraphics[width=0.49\textwidth]{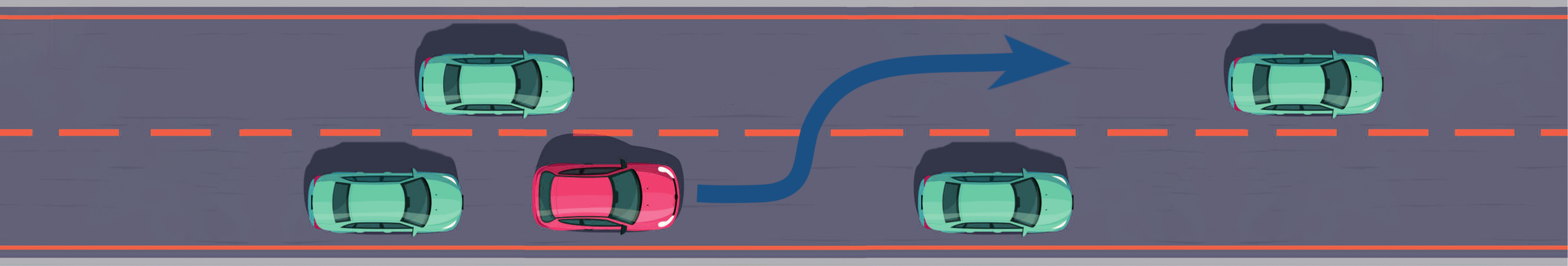}
	\caption{Example of proactive collision avoidance in a highway scenario: if its front vehicle suddenly brakes, the ego vehicle (pink) avoids front and rear-end collision with other road users (green) by safely moving to the left lane.}\label{fig:cars}
\end{figure}

\IEEEPARstart{W}{hile} robust (worst-case) approaches in Model Predictive Control (MPC) synthesis have been used in automated driving to ensure safe motion planning in uncertain dynamic environments~\cite{Gandhi2021,Dixit2020,Ji2017,Batkovic2021}, they can lead to overly-conservative maneuvers~\cite{Chen2022} and eventually fail in reaching the main control objective. For instance, it is recognized that human drivers do not drive according to worst-case considerations: if they did, an urban driver may never merge into its desired lane when considering the worst-case scenario in predicting the behavior of other traffic participants~\cite{Liu2023}, or a highway driver would activate unnecessary emergency braking when considering the worst-case scenario in predicting the behavior of a cut-in vehicle. Arguably, the way human drivers avoid overly-conservative maneuvers is by taking some stochastic metrics into account during the planning. As an example, Fig.~\ref{fig:cars} shows a scenario of proactive collision avoidance: the ego vehicle (pink) is surrounded by other road users (green). If the front vehicle suddenly brakes, a conservative decision would be to decelerate as well to keep the distance. However, this decision could lead to collision with the rear vehicle. It would be much safer in this scenario for the ego vehicle to proactively avoid the collision by moving to the left lane while keeping a safe distance from all the surrounding road users. In summary, proactive collision avoidance can be understood by three key features: swift response to disturbance (i.e.~danger), optimality in terms of safety, and avoiding propagation of hazard to future time steps, which translates into getting out of an emergency situation as fast as possible.

\subsection{Motion Planning Challenges in Different Scenarios}
Stochastic MPC~\cite{Cannon2009} (SMPC) is used in various collision-avoidance applications to generate a reference trajectory within a dynamic environment e.g. for mobile robots~\cite{Zhu2019,Wei2023,Chai2020} or spacing control in vehicle platoons~\cite{Moser2018,Kazemi2018}. In automated driving applications, \cite{Dahl2019} reviews different threat metrics for risk assessment during maneuvers from collision probability to time-to-collision or distance-to-collision between the ego vehicle and other road users. Since the challenges and requirements of stochastic motion planning in an uncertain environment depend on the driving scenario, two cases can be distinguished: urban or highway. 

In urban driving, the vehicles drive at lower speeds, which allows using kinematic models for the ego vehicle~\cite{Paden2016}. In addition, vehicles can decide among different actions such as turning to different streets at a junction, stopping to park, or merging into another lane~\cite{Liu2020}. Moreover, there is a variety of traffic participants from pedestrians and bicycles to different drivers with their own driving styles that significantly affect the decision-making outcome~\cite{Liu2023}. Therefore, the prediction of other participants should be more comprehensive and intention-aware, and the research in this area has been focusing on robust estimation of feasible space~\cite{Sanchez2022,Bujarbaruah2021} and tractable MPC formulations in the presence of uncertainty in the behavior of other traffic participants~\cite{Zhang2021,Batkovic2023}. 

Conversely, the planning problem in highway scenarios faces two entangled challenges: ensuring that the generated trajectory can be attained and ensuring that solving the planning problem is computationally efficient. On highways, an emergency maneuver at high speed would push the vehicle in the nonlinear regime. In this sense, ensuring that the generated reference trajectory can be attained requires considering the nonlinear tire dynamics within the ego vehicle model for a more accurate prediction of the available tire forces~\cite{Bellegarda2022}. A common solution to avoid unattainable generated trajectories is to design an integrated planner/tracker incorporating a higher-fidelity prediction model of the ego vehicle, e.g.\cite{Laurense2022} proposes serially-cascaded models to allow using different sampling times and prediction horizons for the planning and tracking sub-problems.  However, this technique is applicable to less-aggressive maneuvers only, since both prediction models for the planning and tracking sub-problems are simple. In this sense, hierarchical control design is still the most popular choice in the literature for emergency collision avoidance in highway driving~\cite{Fors2021,Hajiloo2021}, and the kinematic single-track model is often selected as the ego vehicle prediction model~\cite{Hajiloo2021,Brudigam2023,Benciolini2023}. On the other hand, incorporating nonlinear tire dynamics significantly increases the computational complexity of the MPC planning problem, which may prevent a proactive response to danger.

\subsection{Sources of Uncertainty in Highway Driving}

In the highway collision avoidance literature, the stochasticity of the uncertain environment is expressed via chance constraints in the SMPC planning problem. After observing their initial position and velocity, the behavior of the obstacles is forecasted over a prediction horizon by considering a linear, often point mass, model~\cite{Brudigam2023}. Stochastic behavior of obstacles is then modeled by random variables in their prediction model such as their velocity~\cite{Malone2017,Petrovic2022} or acceleration~\cite{Zhu2019}. Sometimes, randomness in the lane change decision is considered as well~\cite{Benciolini2023}. In this sense,~\cite{Schuurmans2023} expresses the trajectory of obstacles using a Markov jump system description, whereas~\cite{Malone2017} uses a hybrid obstacle model including stochastic switching decision between continuing along a straight path or following an arc trajectory. This uncertainty is then propagated over the prediction horizon e.g. by chaos-based approaches~\cite{Nakka2023} or state updates via Kalman filter~\cite{Zhu2019,Brudigam2023}, leading to chance constraints in the SMPC problem. The reference trajectory is found by minimizing a cost function which in the literature has been mainly defined as a convex (often quadratic) function of the states and inputs~\cite{Paternain2020}, such as the velocity-tracking error~\cite{Brudigam2023,Schuurmans2023} to enforce maintaining a constant longitudinal velocity. Unfortunately, in emergency maneuvers and hazardous scenarios, minimizing the probability of collision is more important than tracking errors. In this sense, \cite{Wang2022} uses a potential field function for collision avoidance, but the obstacle behavior is not stochastic. However, the objective function of avoiding collision may have no closed form, such as in~\cite{Malone2017}, due to the stochasticity of the switching decision. There, the objective function is constructed iteratively via reachable sets. 

\subsection{Computational Efficiency in Emergencies}

At the same time, tractability is also crucial and must be traded with the accuracy of the model. For instance, in~\cite{Pcolka2018} a hybrid nonlinear prediction model is considered for the ego vehicle and the exponential growth in computational complexity is compensated by adapting the prediction horizon accordingly. Further, \cite{Schafer2023} suggests successive convexification to improve the initial guess for the nonlinear MPC problem to reduce the number of iterations and \cite{Eiras2022} uses a Mixed-Integer Linear Program (MILP) to find the feasible region and feed it into the nonlinear planning problem to find the optimal trajectory. To the best of our knowledge, no research has been done incorporating tire force dynamics for real-time emergency motion planning in highway scenarios, i.e. fast online solution of the planning optimization problem, while minimizing the probability of collision which leads to a highly- nonlinear formulation for the SMPC problem.

Hybrid modeling frameworks~\cite{Lunze2009} such as the Max-Min-Plus-Scaling (MMPS) formalism~\cite{DeSchutter2020}, are effective tools to reduce the computational complexity of the planning problem while incorporating the nonlinear behavior regime. In this sense, hybridization refers to the approximation of a nonlinear function, e.g.~the prediction model, using a hybrid systems modeling framework. In case of a nonlinear control optimization problem, hybridization can lead to an MILP formulation of the problem that is computationally more efficient to solve, compared to a NonLinear Program (NLP). Sequential Quadratic Programming (SQP) and real-time iteration scheme have been used in the literature where the nonlinear dynamics is linearized at each time step{~\cite{Gros2020}}. However, that approach has limited capability to adequately capture the complexity of the nonlinear behavior along the prediction horizon. The fact that MILPs can be solved to global optimality in a finite number of iterations~\cite{Caregnato-Neto2023} makes them a suitable candidate to formulate the MPC planning problem. 

\subsection{Contributions of this Paper}

In this paper, we propose an SMPC motion planner for emergency collision avoidance in highway scenarios. We present a proactive planner design by minimizing the collision risk as well as improving safety using chance constraints in the SMPC formulation. To avoid generating unattainable trajectories, we incorporate nonlinear tire dynamics (accounting for the nonlinear tire behavior  close to saturation limits) within the prediction model for the ego vehicle and we use MMPS approximation to reduce the computational complexity of the planning problem. As a result, the novelties in our work are twofold:
\begin{enumerate}
	\item introducing the idea of MMPS approximation of the nonlinearities for real-time implementation, and
	\item combining hybridized risk minimization within a stochastic MPC framework for highway path planning.
\end{enumerate}
Moreover, we provide a comprehensive analysis of how various formulations of the MPC planner influence the conservatism and efficiency of the algorithm to proactively avoid a collision in hazardous scenarios and we compare our proposed approach to a method inspired by the state-of-the-art SMPC planner in~\cite{Brudigam2023} during various cases studies.
To verify that the generated trajectories can be attained by our proposed SMPC planner, we simulate the maneuvers using a high-fidelity vehicle model in IPG CarMaker\cite{IPG}.

The paper is structured as follows: Section~\ref{sec:prob} describes the formulation of the predictive planning problem. Section~\ref{sec:mmps} briefly covers the MMPS approximation, and Section~\ref{sec:reprob} explains our approach in reformulating and solving the SMPC problem. Simulation results and comparisons to the state-of-the-art SMPC planner and the built-in collision avoidance module in IPG CarMaker are presented in Section~\ref{sec:sim}.  Finally, we conclude this paper in Section~\ref{sec:conc}.

The notation is this work is rather standard. The state and input vectors at time step $k$ are represented by $s (k)$ and $u (k)$, respectively. We use a tilde symbol, e.g.~as in $\tilde{s} (k)$, to denote the trajectory of a signal along the prediction horizon. The probability is expressed by the $\Pr$ symbol.


\section{Problem Formulation}\label{sec:prob}

Given a predicted state trajectory $\tilde{s}$ at control time step $k$ along the next $\Np$ steps as
\begin{align}
	& \tilde{s} (k) = \begin{bmatrix} s^T (k+1 | k) & \dots & s^T(k+\Np | k)\end{bmatrix}^T,\label{eq:stilde}
\end{align}
the SMPC planning optimization problem can be formulated by the generic form
\begin{subequations}\label{eq:prob}
\begin{align}
	\min_{\tilde{s} (k),\tilde{u}(k)} & J (\tilde{s} (k)) , \label{eq:probj} \\
	\text{s.t.} \;\; & s(k+ i | k) = f (s(k + i -1 | k),u(k + i -1)), \label{eq:probc1}\\
	& g \left(s(k+ i -1 | k),u(k+i - 1)\right) \leqslant 0,\label{eq:probc3}\\
	& \Pr (s(k+ i | k) \in \mathcal{S}_k) \geqslant 1-\epsilon, \label{eq:probc2}\\
	& \hspace{0.22\textwidth} \forall i \in \{1, \dots, \Np\}, \notag
\end{align}
where $J$ represents the cost function, usually formulated as deviations from a desired velocity or divergence from a globally-planned reference trajectory. Further, the planning problem is constrained to the prediction model of the ego vehicle $f(\cdot)$ via (\ref{eq:probc1}), general nonlinear constraints $g(\cdot)$ (\ref{eq:probc3}), and the chance constraints (\ref{eq:probc2}) where $\mathcal{S}_k$ is the safe or confidence region in step $k$ and $\epsilon$ is the minimum acceptable probability for constraint violation and is selected to be close to 0. Based on the requirements for highway emergencies, $J$, $f$, $g$ and $\mathcal{S}_k$, often need to be selected in such a way that (\ref{eq:prob}) would be an NLP, hence computationally expensive to solve in real time. As explained in Section~\ref{sec:intro}, we use MMPS approximation of the nonlinearities to facilitate obtaining an MILP reformulation of (\ref{eq:prob}) and to improve the computational efficiency. This is further discussed in the next section.
\end{subequations}


\section{MMPS Approximation}\label{sec:mmps}

As the name suggests, MMPS systems are modeled using max, min, plus, and scaling operators and are equivalent to continuous piecewise-affine systems~\cite{Lunze2009}. Any MMPS function $f\mmps$ can be described by either a conjunctive or a disjunctive canonical form~\cite{DeSchutter2004}:
\begin{subequations}\label{eq:mmpscanon}
	\begin{align}
		&f_{\rm{con}} (\chi) = \min_{p = 1, \dots, P} \max_{q = 1, \dots, m_p} \left(\gamma_{p,q}^T \chi + \nu_{p,q}\right), \label{eq:minmax}\\
		&f_{\rm{dis}} (\chi) = \max_{q = 1, \dots, Q} \min_{p = 1, \dots, n_q} \left(\phi_{p,q}^T \chi + \omega_{p,q}\right), \label{eq:maxmin}
	\end{align}
\end{subequations}
where $\gamma$ and $\phi$ are vectors, $\nu$ and $\omega$ are scalars, and $P$, $Q$, $m_p$, and $n_q$ are integers determining the number of nested $\min$ and $\max$ operators.

A nonlinear (scalar) function $f: \mathcal{D} \to \mathbb{R}$ can be approximated by an MMPS form $[f]\mmps$ in compact state domain $\mathcal{D}$ via solving the nonlinear optimization problem 
\begin{align}
	\min_{\mathcal{A}} \int\limits_{\mathcal{D}} \dfrac{\left\Vert f(\chi) - [f]\mmps(\chi) \right\Vert_2}{\left\Vert f(\chi) \right\Vert_2 + \epsilon_0} \; d \chi,
	\label{eq:fapprox}
\end{align}
where $[ . ]\mmps$ represents the MMPS approximation of the corresponding argument with either forms in (\ref{eq:mmpscanon}) and $\mathcal{A}$ collects the decision variables for fixed values of $P$, $Q$, $m_p$, and $n_q$ as
\begin{align}
	\mathcal{A} = \begin{cases}
		\left(\left\{ \gamma_{p,q} \right\}, \left\{ \nu_{p,q} \right\} \right)_{\substack{p = 1,\dots,P\\q=1,\dots,m_p}} & \rm{if} \; [ \; \emph{f} \; ]\mmps = \emph{f}_{\rm{con}}\\
		\left(\left\{ \phi_{p,q} \right\}, \left\{ \omega_{p,q} \right\} \right)_{\substack{q = 1,\dots,Q\\p=1,\dots,n_q}} & \rm{if} \; [ \; \emph{f} \; ]\mmps = \emph{f}_{\rm{dis}}\\
	\end{cases}.\label{eq:decdef}
\end{align}
Note that $\mathcal{A}$ is a tuple of vector and scalar sets since it is necessary to preserve their order in the MMPS forms. The positive value $\epsilon_0>0$ added to the denominator in (\ref{eq:fapprox}) serves to avoid division by very small values for $\Vert f(\chi) \Vert_2 \approx 0$.

In the next steps, we hybridize a suitable nonlinear prediction model for the ego vehicle by solving (\ref{eq:fapprox}) for the nonlinear terms within the vehicle model and use our information of the shape and form of each nonlinearities to select their respective approximation forms in (\ref{eq:mmpscanon}) and the values of the integer pairs $(P, m_p)$ or $(Q, n_q)$. Problem ({\ref{eq:fapprox}}) is a smooth NLP which can be solved by e.g.\ sequential quadratic programming and multi-start strategy.


\section{Problem Reformulation and Solution Approach}\label{sec:reprob}

\subsection{Obstacle Vehicle Model}
Given $\No$ obstacles on the road, the states of the $\eta$-th obstacle where $\eta \in \{1, \dots, \No\}$ at time step $k$ are expressed by the stochastic vector $z^{(\eta)} (k)$ defined as
\begin{align}
	& z^{(\eta)} (k) = \begin{bmatrix} x\obs^{(\eta)} (k) & y\obs^{(\eta)} (k) & \dot{x}\obs^{(\eta)} (k) & \dot{y}\obs^{(\eta)} (k) \end{bmatrix}^T,\label{eq:zdef}
\end{align}
with the Gaussian distribution
\begin{align}
	& z^{(\eta)} (k) \sim \mathcal{N} \left(\xi^{(\eta)}  (k) ,\Xi^{(\eta)}  (k) \right),\label{eq:obsgaus}
\end{align}
where $\xi$ and $\Xi$ respectively indicate the mean vector and the covariance matrix as
\begin{align}
	& \xi^{(\eta)} (k) = \begin{bmatrix} \xi_x^{(\eta)} (k) & \xi_y^{(\eta)} (k) & \xi_{\dot{x}}^{(\eta)} (k) & \xi_{\dot{y}}^{(\eta)} (k) \end{bmatrix}^T,\\
	& \Xi^{(\eta)} (k) = \begin{bmatrix} \sigma_x^{(\eta)} (k) & 0 & 0 & 0 \\ 0 & \sigma_y^{(\eta)} (k) & 0 & 0 \\ 0 & 0 & \sigma_{\dot{x}}^{(\eta)} (k) & 0 \\ 0 & 0 & 0 & \sigma_{\dot{y}}^{(\eta)} (k) \end{bmatrix}.\label{eq:zgausdef}
\end{align}
\begin{figure}[hbt]\centering
	\includegraphics[width=0.48\textwidth]{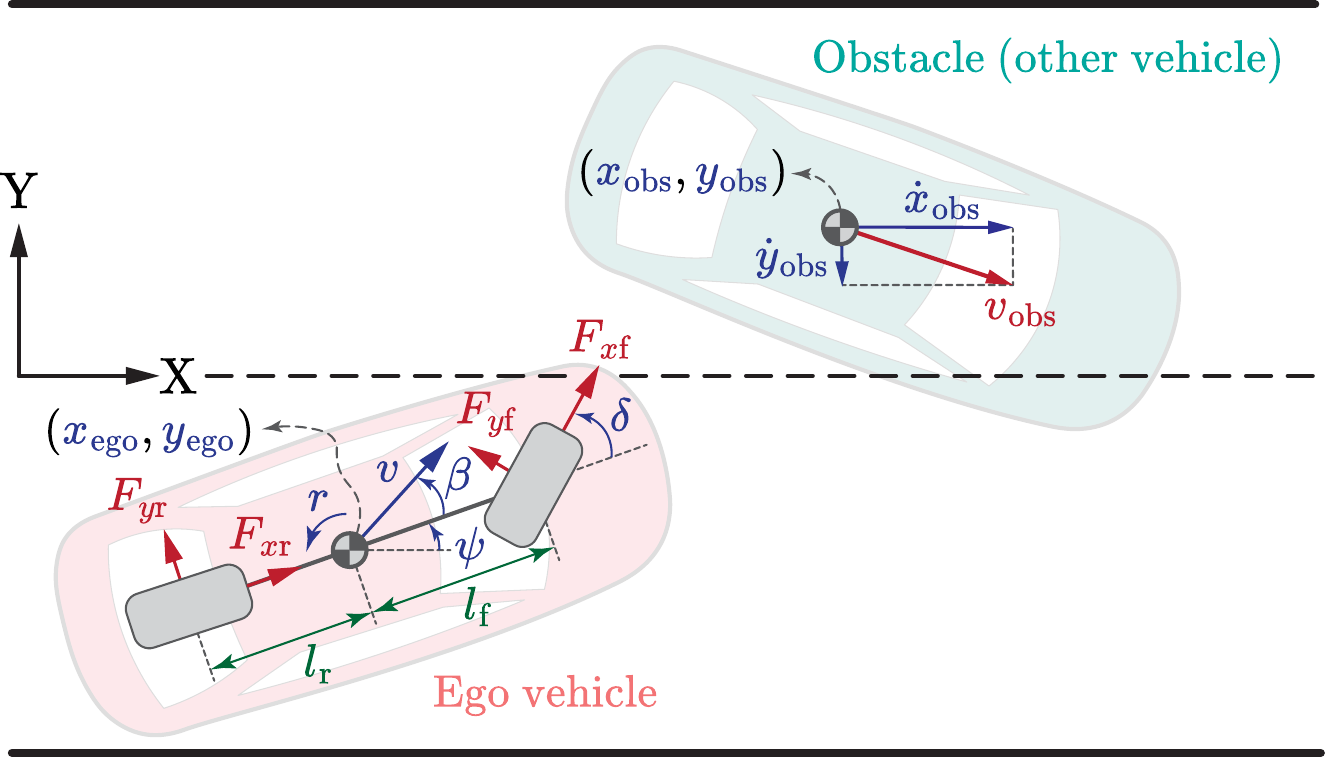} 
	\caption{Model configuration for the ego vehicle and the obstacles on the road.}\label{fig:models} 
\end{figure}
\begin{rem}
	We use discretized double integrator dynamics to model the obstacle behavior and update variance and mean using Kalman predictions. Note that the actual covariance matrix does not remain diagonal, but it is customary to consider a reduced or approximated covariance matrix including the diagonal elements of $\Xi$ associated with the target states~\cite{Chen2020,Kuo2021,Tiger2021,Brudigam2022-ext} for computational efficiency; an approach we use in this paper as well.
\end{rem}
More specifically, we use a point mass model~\cite{Brudigam2023} for the obstacles in Fig.~\ref{fig:models}, expressed by
\begin{subequations}
	\begin{align}
		& z^{(\eta)}  (k+1) = A z^{(\eta)} (k) + B w^{(\eta)}  (k) + \nu^{(\eta)} (k), \label{eq:obsm1}\\
		& w^{(\eta)}  (k) = K \left(z\rf^{(\eta)} (k) - z^{(\eta)}  (k)\right),\label{eq:obsm2}
	\end{align}\label{eq:obsmodel}
\end{subequations}
where $A$ and $B$ are the state and input matrices resulting from discretized double integrator dynamics, $w$ represents the input signal as
\begin{align}
	& w^{(\eta)} (k) = \begin{bmatrix} \ddot{x}\obs^{(\eta)} (k) & \ddot{y}\obs^{(\eta)} (k)  \end{bmatrix}^T,\label{eq:udef}
\end{align}
$\nu \sim \mathcal{N} (0_{4 \times 1}, \Xi_0)$ represents the process noise, and $K$ is a stabilizing gain such that the obstacle tracks its corresponding reference state $z\rf$. Based on the current state of the obstacles, we assume that the obstacles intend to keep their current longitudinal velocity and their lateral position on the road. Therefore, $z\rf$ is heuristically estimated in the planning layer at each time step based on the most likely/expected behavior of the other road users. The covariance matrix for each obstacle is updated at each time step in line with Kalman prediction by
\begin{subequations}
	\begin{align}
		\xi^{(\eta)} (k+1) = (A - B K) \xi^{(\eta)} (k) + B K z\rf^{(\eta)} (k),\\
		\Xi^{(\eta)} (k+1) = (A - B K) \Xi^{(\eta)} (k) (A - B K)^T + \Xi^{(\eta)}_0.
	\end{align}\label{eq:obscov}
\end{subequations}
with $\Xi_0$ being the initial estimate of the covariance matrix of the process noise. Using the Gaussian distribution in (\ref{eq:obsgaus}), we define $p_k^{(\eta)}$ to express the probability density function for the presence of obstacle $\eta \in \{1, \dots, \No\}$ on the road as
\begin{multline}
	p_k^{(\eta)} (x,y) = \\
	\dfrac{\exp\left(-\left(\dfrac{x -\xi_x^{(\eta)}(k)}{ {\sqrt{2}\sigma^{(\eta)}_x}(k)}\right)^2-\left(\dfrac{y -\xi_y^{(\eta)}(k)}{ {\sqrt{2} \sigma^{(\eta)}_y}(k)}\right)^2\right)}{2 \pi \sigma^{(\eta)}_x (k) \sigma^{(\eta)}_y (k)},\label{eq:pdfbeta}
\end{multline}
which is used to develop the probability function $\mathbb{P}$ for the state vector $s (k)$ defined in (16) using a chi-squared distribution (see~\cite{Brudigam2023}) and taking into account the unsafe area $\Omega^{(\eta)}$, as 
\begin{align}
	\mathbb{P}^{(\eta)} (s(k)) = \Pr\left((x\e (k), y\e (k) \in \Omega^{(\eta)})\right).\label{eq:probeta}
\end{align}
The unsafe set $\Omega$ for each obstacle is defined as an area that the center of gravity of the ego vehicle must avoid, and it is an ellipse calculated by considering the position and size of both ego and obstacle vehicles as known parameters~\cite{Dahl2019}.

\subsection{Hybrid Ego Vehicle Model}

The ego vehicle prediction model as shown in Fig.~\ref{fig:models} is described by a dynamic bicycle model~\cite{Stano2023} with a small-angle assumption for $\delta$ (reasonable in highway scenarios~\cite{Dixit2020})  
\begin{subequations}\label{eq:evmodel}
\begin{align}
	&\dot{x}\e = v \cos (\psi + \beta), \label{eq:dxnl}\\
	&\dot{y}\e = v \sin (\psi + \beta), \label{eq:dynl}\\
	&\dot{\psi} = r, \label{eq:dpsinl}\\
	&\dot{v} = \dfrac{1}{m} \left[\Fxf - \Fyf \delta + \Fxr\right]+ v \beta r,\label{eq:dvnl}\\
	&\dot{\beta} = \dfrac{1}{m v} \left[\Fyf + \Fyr\right]- r,\label{eq:dbetanl}\\
	&\dot{r} = \dfrac{1}{I_{zz}} \left[\Fxf \delta \; \lf + \Fyf  \; \lf - \Fyr \;\lr \right],\label{eq:drnl}\\
	&\dot{\delta} = d_\delta,\label{eq:ddeltanl}
\end{align}
\end{subequations}
with $\Fxf$, $\Fxr$, and $d_\delta$ as inputs. All the variables and system parameters are described in Tables~\ref{tab:vars} and~\ref{tab:params}, and the state vector $s$ at time step $k$ is expressed by
\begin{align}&
		s (k) = \label{eq:sdef} \\
	&\begin{bmatrix}
			x\e (k) & y\e(k) & \psi (k) & v(k) & \beta (k) & r (k) & \delta (k)
		\end{bmatrix}^T.\notag
\end{align}

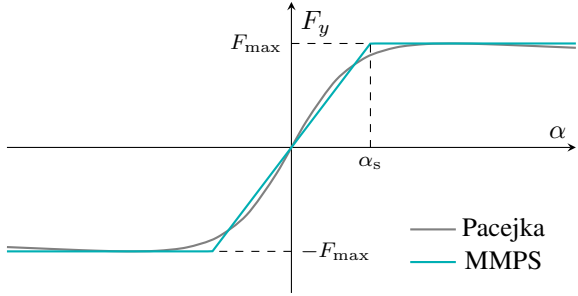
\begin{figure}[hbtp]\centering
\begin{tikzpicture}
\begin{axis}[height=0.3\textwidth,width=0.5\textwidth,ytick=0,xtick=\empty,clip=false,ymajorgrids,
	xlabel=$\alpha$,ylabel=$F_y$,xmin=-0.3,xmax=0.3,ymin=-1.4,ymax=1.4,legend pos=south east,axis x line=middle,axis y line=middle,legend style={draw=none, fill opacity=0.0, text opacity = 1}]
	\addplot[color=gray,smooth,thick,samples=20,domain=-0.3:0.3]{sin(deg(1.45*rad(atan(11.24*x))))};
	\addplot[color=TealBlue,thick,samples=120,domain=-0.3:0.3]{min(max(12*x,-1),1)};
	\legend{Pacejka,MMPS}
	\addplot[color=black, dashed] coordinates {(0,1) (1/12,1) (1/12,0)};
	\addplot[color=black, dashed] coordinates {(0,-1) (-1/12,-1)};
	\node[left] at (0.0,1) {\footnotesize$F_{\rm{max}}$};
	\node[right] at (0.0,-1) {\footnotesize$-F_{\rm{max}}$};
	\node[below] at (1/12,0) {\footnotesize$\als$};
\end{axis}
\end{tikzpicture}
\caption{Pacejka tire model and its MMPS approximation}\label{fig:fytire}
\end{figure}

The tire forces should satisfy the tire saturation limits 
\begin{subequations}\label{eq:nlkamm}
\begin{align}
	\Fxf^2 + \Fyf^2 \leqslant \left(\mu \Fzf\right)^2,\\
	\Fxr^2 + \Fyr^2 \leqslant \left(\mu \Fzr\right)^2,
\end{align}
\end{subequations}
also known as Kamm circle constraint~\cite{Rajamani2011}. Considering the slip angles
\begin{subequations}\label{eq:alphas}
\begin{align}
	&\alf = \delta - \beta + \dfrac{\lf r}{v}, \label{eq:alf}\\
	&\alr = \dfrac{\lr r}{v} - \beta, \label{eq:alr}
\end{align}
\end{subequations}
we describe the lateral tire forces by MMPS approximations of the Pacejka tire model~\cite{Pacejka2005} shown in Fig.~\ref{fig:fytire} as
\begin{align}
&[F_y]\mmps= F_{\rm{max}} \min \left( \max \left(\dfrac{\alpha}{\als},-1 \right),  1\right),\label{eq:fygen}
\end{align}
where the nonlinear function representing the tire forces on the front and rear axles is approximated by a parametric MMPS function where $F_{\rm{max}}$ and $\als$ respectively correspond to the maximum tire force and the saturation slip angle. 

Substituting the front and rear slip angles in (\ref{eq:fygen}) gives the front and rear lateral tire forces as
\begin{subequations}\label{eq:fymmps}
\begin{align}
	&[\Fyf]\mmps = F_{\rm{max}} \min \left( \max \left(\dfrac{\delta}{\als} - \dfrac{\beta}{\als} + \dfrac{\lf \; r}{\als v_0},-1 \right),  1\right),\\
	&[\Fyr ]\mmps= F_{\rm{max}} \min \left( \max \left(\dfrac{\lr}{\als v_0} r  - \dfrac{1}{\als} \beta,-1 \right),  1\right).
\end{align}
\end{subequations}
Using the MMPS approximation of the other nonlinear terms in the ego vehicle model, we obtain an MMPS formulation for the ego vehicle model expressed by
\begin{subequations}\label{eq:evmmps}
\begin{align}
	&\dot{x}\e =\max\{v, v_0 \; [\cos (\psi + \beta)]\mmps\}, \label{eq:dx}\\
	&\dot{y}\e = v_0 [\sin (\psi + \beta)]\mmps, \label{eq:dy}\\
	&\dot{\psi} = r, \label{eq:dpsi}\\
	&\dot{v} = \dfrac{\Fxf+\Fxr}{m} - \dfrac{[\delta \; \Fyf]\mmps}{m} + v_0 \; [\beta r]\mmps,\label{eq:dv}\\
	&\dot{\beta} =  \dfrac{[\Fyf + \Fyr]\mmps}{m v_0}- r,\label{eq:dbeta}\\
	&\dot{r} = \dfrac{\lf \; \delta_0 \; \Fxf}{I_{zz}} + \dfrac{\lf \; [\Fyf ]\mmps}{I_{zz}} - \dfrac{\lr \; [\Fyr]\mmps }{I_{zz}},\label{eq:dr}\\
	&\dot{\delta} = d_\delta,\label{eq:ddelta}	
\end{align}
\end{subequations}
Figure{~\ref{fig:mmps} presents three examples of the nonlinear terms vs.~their MMPS approximations. To find these formulations, we have used information on the form of the nonlinear function and we have selected the number of $\max$ and $\min$ operators accordingly. For instance, in Fig.{~\ref{fig:mmpsdx}}, we use three hyperplanes and two $\max$ and $\min$ operators based on the cosinusoidal shape of the nonlinear function.
\begin{rem}
	Considering the orders of magnitude of variations of the longitudinal velocity over the prediction horizon, the velocity $v$ in (\ref{eq:dynl}), (\ref{eq:dvnl}) and (\ref{eq:dbetanl}) can be approximated as a fixed parameter over the prediction horizon and can be taken equal to the current measured velocity. Moreover, in cases where $v$ is multiplied by cosine terms with values close to 1, we take the maximum value between the velocity $v$ and the MMPS approximation with $v=v_0$ in (\ref{eq:dx}) to ensure the inclusion of numerically significant effects resulting from variations in $v$ when \mbox{$\psi+\beta\approx0$} in (\ref{eq:dx}). A similar approach is used for $\delta$ in (\ref{eq:drnl}) where its variations are included in the MMPS tire forces and the current steering angle is used as a parametric coefficient for the first term. 
\end{rem}
\begin{rem}
	After MMPS approximation of the continuous-time model of the ego vehicle, (\ref{eq:evmmps}) can be discretized e.g.\ using forward Euler method and a proper sampling time to be incorporated in the SMPC formulation in (\ref{eq:prob}).
\end{rem}

Further, the Kamm circle constraints in (\ref{eq:nlkamm}) are approximated using MMPS function in Fig.~\ref{fig:mmpskammf}. Note that due to different ranges of $\Fxf$ and $\Fxr$, the front and rear force magnitudes are approximated by the maximum of respectively three and four affine functions, to appropriately capture the form of the nonlinear function. The maximum tire forces on the front and rear axles are functions of the online measurements of the friction coefficient $\mu$, which we assume available via a friction estimator~\cite{Laurense2022,Stano2023}, as
\begin{equation*}
F_{\rm{max}} = \min \{\mu \Fzf , \mu \Fzr\}.
\end{equation*}
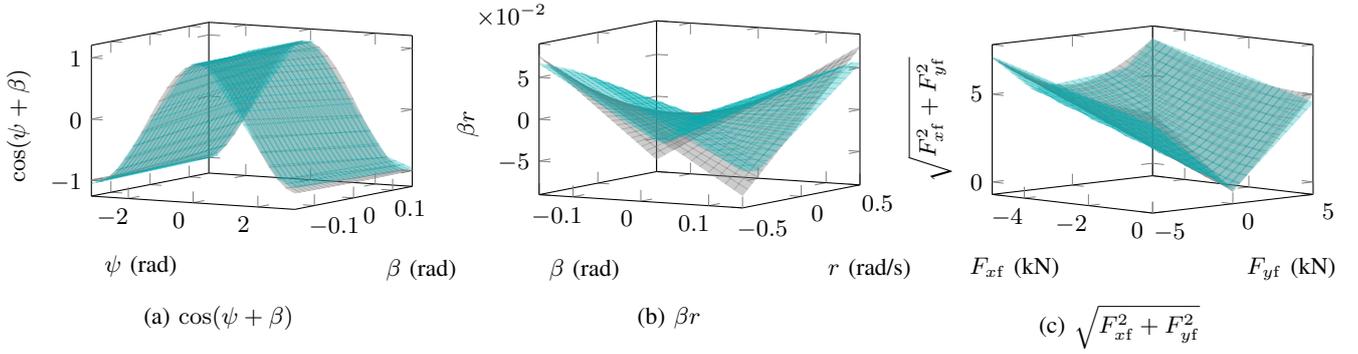
\begin{figure*}[hbtp]
\begin{center}
\begin{subfigure}[t]{0.32\linewidth}\centering
\begin{tikzpicture}
\begin{axis}[height=0.7\textwidth,width=\textwidth,colormap/blackwhite,view={30}{10},
	xlabel=$\psi$ (rad),ylabel=$\beta$ (rad),zlabel=$\cos (\psi+\beta)$,
	legend style={at={(0.5,1.3)}, anchor=north,draw=none, fill opacity=0.0, text opacity = 1},
	xmin=-pi,xmax=pi,ymin=-0.15,ymax=0.15,legend columns=-1]
	\addplot3[surf,shader=flat,gray,domain=-pi:pi,domain y=-0.15:0.15,opacity=0.4]
	{cos(deg(x)+deg(y))};
	\addplot3[surf,shader=flat,TealBlue,domain=-pi:pi,domain y=-0.15:0.15,opacity=0.3]
	{min(max(-0.2*x-0.2*y-0.3,-0.8*x-0.8*y+1.2),max(0.8*x+0.8*y+1.2,0.2*x+0.15*y-0.4),1)};
\end{axis}
\end{tikzpicture}
\subcaption{$\cos (\psi+\beta)$}\label{fig:mmpsdx}
\end{subfigure}			
\begin{subfigure}[t]{0.32\linewidth}
\begin{tikzpicture}
\begin{axis}[height=0.7\textwidth,width=\textwidth,colormap/blackwhite,view={30}{10},
	xlabel=$\beta$ (rad),ylabel=$r$ (rad/s),zlabel=$\beta r$,
	legend style={at={(0.5,1.3)}, anchor=north,draw=none, fill opacity=0.0, text opacity=1},
	ymin=-0.5,ymax=0.5,xmin=-0.15,xmax=0.15,legend columns=-1]
	\addplot3[surf,shader=flat,gray,domain=-0.15:0.15,domain y=-0.5:0.5,opacity=0.4]
	{x*y};
	\addplot3[surf,shader=flat,TealBlue,domain=-0.15:0.15,domain y=-0.5:0.5,opacity=0.3]
	{max(min(-0.44*x-0.02*y+0.01,-0.11*x-0.12*y-0.01),min(0.40*x-0.0*y-0,-0.03*x+0.12*y))};
\end{axis}
\end{tikzpicture}
\subcaption{$\beta r$}\label{fig:mmpsbr}
\end{subfigure}	
\begin{subfigure}[t]{0.32\linewidth}
\begin{tikzpicture}
\begin{axis}[height=0.7\textwidth,width=\textwidth,colormap/blackwhite,view={45}{10},
	xlabel=$\Fxf$ (kN),ylabel=$\Fyf$ (kN),zlabel=$\sqrt{\Fxf^2+\Fyf^2}$,
	legend style={at={(0.5,1.3)}, anchor=north,draw=none, fill opacity=0.0, text opacity=1},
	ymin=-5,ymax=5,xmin=-5,xmax=0,legend columns=-1]
	\addplot3[surf,shader=flat,gray,domain=-5:0,domain y=-5:5,opacity=0.4]
	{(x^2+y^2)^0.5};
	\addplot3[surf,shader=flat,TealBlue,domain=-5:0,domain y=-5:5,opacity=0.3]
	{max(-0.56*x-0.83*y+0.16,-1.05*x+0.10*y-0.02,-0.51*x+0.89*y+0.07)};
\end{axis}
\end{tikzpicture}
\subcaption{$\sqrt{\Fxf^2+\Fyf^2}$}\label{fig:mmpskammf}
\end{subfigure}	
\caption{Plots of example nonlinear terms in the ego vehicle prediction model and their MMPS approximations}\label{fig:mmps}
\end{center}
\end{figure*}
\begin{table}[htbp]\caption{System variables and their bounds in the case study}
\begin{center}
\begin{tabular}{c|c|c|c}
	\hline
	\textbf{Var.} & \textbf{Definition} & \textbf{Unit} & \textbf{Bounds}\\
	\hline
	$x\e$ & Longitudinal position of the ego vehicle & m & [0, $\infty$]\\
	$x\obs$ & Longitudinal position of the obstacle & m & [0, $\infty$]\\
	$\dot{x}\obs$ & Longitudinal velocity of the obstacle & m/s & [5, 50]\\
	$y\e$ & Lateral position of the ego vehicle & m & [-6, 6]\\	
	$y\obs$ & Lateral position of the obstacle & m & [-6, 6]\\
	$\dot{y}\obs$ & Lateral velocity of the obstacle & m/s & [-5,5]\\
	$v$ & Velocity of the ego vehicle & m/s & [5, 50]\\
	$\beta$ & Sideslip angle & rad & [-0.2,0.2]\\
	$\psi$ & Yaw angle & rad & [$-\pi$,$\pi$] \\
	$r$ & Yaw rate & rad/s & [-0.5, 0.5]\\
	$\delta$ & Steering angle (road)& rad & [-0.2, 0.2]\\
	$\Fxf$ & Longitudinal force on the front axis & N & [-5000, 0]\\
	$\Fxr$ & Longitudinal force on the rear axis & N & [-5000, 5000]\\
	$\Fyf$ & Lateral force on the front axis & N & --\\
	$\Fyr$ & Lateral force on the rear axis & N & --\\
	$\Fzf$ & Normal load on the front axis & N & --\\
	$\Fzr$ & Normal load on the rear axis & N & --\\
	$\alf$ & Front slip angle & rad & --\\
	$\alr$ & Rear slip angle & rad & --\\
	\hline
\end{tabular}\label{tab:vars}
\end{center}
\end{table}
\begin{table}[htbp]\caption{System parameters}
\begin{center}
\begin{tabular}{c|c|c|c}
	\hline
	\textbf{Par.} & \textbf{Definition} & \textbf{Value} & \textbf{Unit}\\
	\hline
	\multicolumn{4}{c}{\emph{Fixed Parameters}}\\
	\multicolumn{4}{c}{\emph{(IPG CarMaker BMW vehicle model)}}\\
	\hline
	$m$ & Vehicle mass & 1970 & kg\\
	$I_{zz}$ & Inertia moment about z-axis & 3498& kg/m$^2$\\
	$\lf$ & CoG$^{\ast}$ to front axis distance& 1.4778 & m\\
	$\lr$ & CoG to rear axis distance & 1.4102 & m\\
	$\Fzf$ & Normal load on the front axis & 7926 & N \\
	$\Fzr$ & Normal load on the rear axis & 8303 & N\\
	$\als$ & Saturation slip angle & 0.09 & rad \\
	$t_{\rm{s}}$ & Planner sampling time & 0.2 & s \\
	$\Np$ & Prediction horizon & 10 & -- \\
	\hline
	\multicolumn{4}{c}{\emph{Varying Parameters}}\\
	\multicolumn{4}{c}{\emph{(Measured Online)}}\\
	\hline
	$F_{\rm{max}}$ & Maximum tire force & -- & N \\
	$\mu$ & Friction coefficient & -- & --\\
	$s_0$ & Initial EV state vector & -- & -- \\
	$x\rf$ & Globally-planned reference trajectory & -- & -- \\
	$v_0$ & Initial velocity & -- & m/s \\
	$\delta_0$ & Initial steering angle & -- & rad \\
	\hline
	\multicolumn{4}{l}{$^{\ast}$Center of Gravity}
\end{tabular}\label{tab:params}
\end{center}
\end{table}

\subsection{Chance Constraints and Collision Risk Function}
To hybridize the probability function $\mathbb{P}$ in (\ref{eq:probeta}), we approximate it by the MMPS function $[\mathbb{P}]\mmps$ as illustrated in Fig.~\ref{fig:prob}. The MMPS approximation $[\mathbb{P}]\mmps$ is a probability function as well and is used as a chance constraint in the SMPC formulation. 
\begin{figure}[hbtp]
\begin{center}
\begin{tikzpicture}
\begin{axis}[
	enlargelimits=false,axis lines=center,view={135}{25},legend pos=north west,legend style={at={(0.0,1)},anchor=north west},
	height=0.35\textwidth,width=0.5\textwidth,xmin=0, xmax=1.3,ymin=0, ymax=1.3,zmin=0, zmax=0.7,
	xtick=\empty, ytick=\empty, ztick=\empty,axis on top,set layers,
	xlabel=$x$, ylabel=$y$, zlabel=$\mathbb{P}$,enlargelimits=upper,clip=false,]
	\begin{pgfonlayer}{axis background}
		\coordinate (L1) at (0.9,0,0);
		\coordinate (L2) at (0.9,-0.4,0);
		\coordinate (L3) at (-0.9,-0.4,0);
		\coordinate (L4) at (-0.9,0.4,0);
		\coordinate (L5) at (0,0.4,0);
		\coordinate (LO) at (0,0,0);
		\coordinate (H1) at (0.08,0,0.4);
		\coordinate (H2) at (0.08,-0.04,0.4);
		\coordinate (H3) at (-0.08,-0.04,0.4);
		\coordinate (H4) at (-0.08,0.04,0.4);
		\coordinate (H5) at (0,0.04,0.4);
		\coordinate (HO) at (0,0,0.4);
		\begin{scope}[canvas is xy plane at z=0]
			\draw[fill=TealBlue,opacity=0.4] (0,0) circle (1cm);
		\end{scope}
		\path[fill=white,opacity=1] (LO)--(0,1.5,0)--(1.5,0,0)--(L1)--(L2)--(L3)--(L4)--(L5)--(LO);
		\addplot3[ultra thin,trig format plots=rad,surf,gray,shader=flat,draw=gray,
		point meta={sqrt(x^2 +y^2)},colormap/blackwhite,opacity=0.4,forget plot,
		domain=0.001:pi/2,unbounded coords=jump,samples=21,y domain=pi/2:2*pi,samples y = 41,
		z buffer=sort,] 
		({sin(x)*cos(y)},{sin(x)*sin(y)},{0.5*exp(-(sin(x)^2*cos(y)^2/0.2+sin(x)^2*sin(y)^2/0.05)});
	\end{pgfonlayer}		
	\begin{pgfonlayer}{axis background}
		\coordinate (HL1) at (0.9,0,0);
		\coordinate (HL2) at (0.9,-0.4,0);
		\coordinate (HL3) at (-0.9,-0.4,0);
		\coordinate (HL4) at (-0.9,0.4,0);
		\coordinate (HL5) at (0,0.4,0);
		\coordinate (HLO) at (0,0,0);
		\coordinate (HH1) at (0.16,0,0.5);
		\coordinate (HH2) at (0.16,-0.08,0.5);
		\coordinate (HH3) at (-0.16,-0.08,0.5);
		\coordinate (HH4) at (-0.16,0.08,0.5);
		\coordinate (HH5) at (0,0.08,0.5);
		\coordinate (HHO) at (0,0,0.5);
		\begin{scope}[canvas is xy plane at z=0]
			\draw[fill=RedViolet,opacity=0.4] (0,0) circle (1cm);
		\end{scope}
		\path[fill=white,opacity=1] (HLO)--(0,1.5,0)--(1.5,0,0)--(HL1)--(HL2)--(HL3)--(HL4)--(HL5)--(HLO);
		\addplot3[ultra thin,trig format plots=rad,surf,gray,shader=flat,draw=gray,
		point meta={sqrt(x^2 +y^2)},colormap/blackwhite,opacity=0.4,forget plot,
		domain=0.001:pi/2,unbounded coords=jump,samples=21,y domain=pi/2:2*pi,samples y = 41,
		z buffer=sort,] 
		({sin(x)*cos(y)},{sin(x)*sin(y)},{0.5*exp(-(sin(x)^2*cos(y)^2/0.15+sin(x)^2*sin(y)^2/0.03)});
	\end{pgfonlayer}		
	\begin{pgfonlayer}{axis foreground}
		\draw[fill=RedViolet,ultra thin,opacity=0.2] (HHO)--(HH1)--(HH2)--(HH3)--(HH4)--(HH5)--(HHO);
		\draw[fill=RedViolet,ultra thin,opacity=0.2] (HH1)--(HL1)--(HL2)--(HH2)--(HH1);
		\draw[fill=RedViolet,ultra thin,opacity=0.2] (HH3)--(HL3)--(HL2)--(HH2)--(HH3);
		\draw[fill=RedViolet,ultra thin,opacity=0.2] (HH3)--(HL3)--(HL4)--(HH4)--(HH3);
		\draw[fill=RedViolet,ultra thin,opacity=0.2] (HH4)--(HL4)--(HL5)--(HH5)--(HH4);		
		\addplot3[very thick,trig format plots=rad,draw=black,
		domain=0.001:pi/2,samples=81,samples y = 1,forget plot,] 
		({sin(x)*cos(0)}, {sin(x)*sin(0)}, {0.5*exp(-(sin(x)^2/0.15)});
		\addplot3[very thick,trig format plots=rad,draw=black,forget plot,
		domain y =0.001:pi/2,samples y =81,samples = 1] 
		({sin(0)*sin(y)}, {sin(y)*cos(0)}, {0.5*exp(-(y^2/0.03))});
		\draw[draw=RedViolet,very thick] (HHO)--(HH1)--(HL1)--(1,0,0);
		\draw[draw=RedViolet,very thick] (HHO)--(HH5)--(HL5)--(0,1,0);
	\end{pgfonlayer}
	\begin{pgfonlayer}{axis foreground}
		\draw[fill=TealBlue,ultra thin,opacity=0.2] (HO)--(H1)--(H2)--(H3)--(H4)--(H5)--(HO);
		\draw[fill=TealBlue,ultra thin,opacity=0.2] (H1)--(L1)--(L2)--(H2)--(H1);
		\draw[fill=TealBlue,ultra thin,opacity=0.2] (H3)--(L3)--(L2)--(H2)--(H3);
		\draw[fill=TealBlue,ultra thin,opacity=0.2] (H3)--(L3)--(L4)--(H4)--(H3);
		\draw[fill=TealBlue,ultra thin,opacity=0.2] (H4)--(L4)--(L5)--(H5)--(H4);		
		\addplot3[very thick,trig format plots=rad,draw=black,
		domain=0.001:pi/2,samples=81,samples y = 1,forget plot,] 
		({sin(x)*cos(0)}, {sin(x)*sin(0)}, {0.5*exp(-(sin(x)^2/0.15)});
		\addplot3[very thick,trig format plots=rad,draw=black,forget plot,
		domain y =0.001:pi/2,samples y =81,samples = 1] 
		({sin(0)*sin(y)}, {sin(y)*cos(0)}, {0.5*exp(-(y^2/0.03))});
		\draw[draw=TealBlue,very thick] (HO)--(H1)--(L1)--(1,0,0);
		\draw[draw=TealBlue,very thick] (HO)--(H5)--(L5)--(0,1,0);
		\addlegendimage{area legend, fill=gray!50};
		\addlegendimage{area legend, fill=TealBlue!50};
		\addlegendimage{area legend, fill=RedViolet!50};
		\legend{$\mathbb{P}$,$[\mathbb{P}]\mmps$,$[\hat{\mathbb{P}}]\mmps$};
	\end{pgfonlayer}
\end{axis}
\end{tikzpicture}
\caption{Conceptual illustration of the Gaussian probability function $\mathbb{P}$, of its MMPS approximation and of the MMPS proxy functions. The approximations are valid in the compact domain $\mathcal{D}$.}\label{fig:prob}
\end{center}
\end{figure}
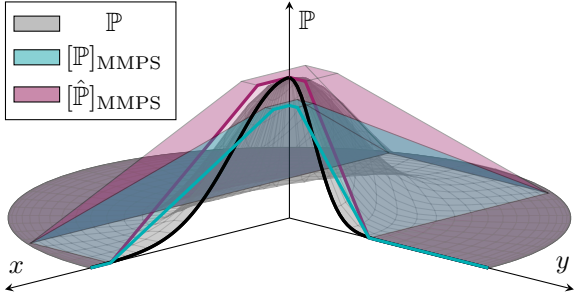

Since the chance constraints must be bounded in such a way that the probability of constraint violation is very low (to improve safety), the accuracy of the MMPS approximation is more important in regions close to $\mathbb{P} = 0$. Therefore, we obtain $[\mathbb{P}]\mmps$ by approximating the Gaussian probability density function (\ref{eq:pdfbeta}) on a compact domain $\mathcal{D}$ defined by the road boundaries in the lateral direction, and the maximum possible longitudinal displacement during the prediction horizon, via solving (\ref{eq:fapprox}) and imposing the constraint 
\[\int_{\mathcal{D}}  [p_k]\mmps^{(\eta)} (x,y) = 1,\]
which gives the parametric form for $[\mathbb{P}]\mmps$ from (\ref{eq:probeta}) as
\begin{equation}
	[\mathbb{P}]\mmps (s(k)) = \max \left( \min_{p = 1, \dots, 5} \left(\phi_p^T s (k) + \omega_{p}\right) , 0\right),\label{eq:parmp}
\end{equation}
with $\phi_p$ being affine functions of $\xi_x (k)$, $\xi_y (k)$, $\sigma_x (k)$ and $\sigma_y (k)$. Similar to $\mathbb{P}$, the MMPS approximation $[\mathbb{P}]\mmps$ is a probability function that is used in the chance constraints.

However, $[\mathbb{P}]\mmps$ under-estimates $\mathbb{P}$ in regions close to the peak of $\mathbb{P}$, which is not desired for deriving the collision risk function. To improve safety, we use the MMPS function $[\hat{\mathbb{P}}]\mmps$ in Fig.~\ref{fig:prob} as a proxy of $[\mathbb{P}]\mmps$ to obtain the risk of collision for each point on the road in the presence of other road users. This time, we find $\hat{\phi}_p$ by approximating $p_k$ via (\ref{eq:fapprox}) constrained to
\[[\hat{p}_k]\mmps^{(\eta)} (x,y) \geqslant p_k (x,y), \qquad \forall (x,y) \in \mathcal{D},\]
which gives the proxy function 
\begin{equation}
	[\hat{\mathbb{P}}]\mmps (s(k)) = \max \left( \min_{p = 1, \dots, 5} \left(\hat{\phi}_p^T s (k) + \hat{\omega}_{p}\right) , 0\right),\label{eq:parproxy}
\end{equation}
serving as an over-estimation of $\mathbb{P}$ based on $[\hat{p}_k]\mmps$. Since $[\hat{p}_k]\mmps$ is not a probability density function, $[\hat{\mathbb{P}}]\mmps$ is only used to calculate the risk as the cost and does not serve as an approximation of the probability in evaluating the chance constraint. This separation allows to avoid conservatism in $[\mathbb{P}]\mmps$ within the constraints while seeking safer trajectories by minimizing the over-approximation $[\hat{\mathbb{P}}]\mmps$.

For each time step, the collision risk depends on the probability of the presence of other road users in $(x,y)$. Therefore, the collision risk of $\tilde{s} (k)$ can be defined as
\begin{align}
	& P (\tilde{s} (k)) = \dfrac{1}{\Np}\sum_{i=1}^{\Np} \max_{\eta = \{1, \dots, \No\}} [\hat{\mathbb{P}}]\mmps^{(\eta)} (s (k+i | k)).\label{eq:probfull}
\end{align}
\begin{rem}
	The $\max$ operator in (\ref{eq:probfull}) can be replaced by a sum across the presence probability of all the $\No$ road users. However, this sum may result in a more conservative estimation of the collision risk\footnote{The same argument can be deduced using Boole's inequality.}. For instance, if there are two obstacles with a safe corridor in between where \mbox{$\mathbb{P}^{(1)} = \mathbb{P}^{(2)} = \rho$}, the sum would give a risk of $\approx 2 \rho$ for this area, whereas in a real situation, the chance of two vehicles getting closer is low; furthermore, the real presence probability for both obstacles would be even lower than $\rho$ which is an estimate that does not take into account the effect of the presence of one obstacle on the decisions of other road users.
\end{rem}

\subsection{SMPC Optimization Problem}

We incorporate the presence probability of obstacles into the MPC planner in two ways: first, we ensure a very low probability for the collision by constraining $[\mathbb{P}]\mmps^{(\eta)}$ to be less than a small threshold $\epsilon >0$. Secondly, we minimize the collision risk function $P$ from (\ref{eq:probfull}) in the objective function to not only ensure this safety level, but also to converge to the safest attainable trajectory and to prevent getting close to high-risk areas in a predictive manner. This in fact will lead to a more proactive response to danger during a hazardous scenario, which will be illustrated in an example case later.

The stochastic MPC motion planner is formulated as follows: given a globally-planned reference velocity profile $\tilde{v}\rf$ and the initial states $s_0$, we find the optimal trajectory $\tilde{s}\rf$ by solving
\begin{subequations}\label{eq:smpc}
	\begin{align}
		& \min_{\tilde{s},\tilde{u}} && P (\tilde{s}) + \lambda_{\rm{v}} \Vert \tilde{v} - \tilde{v}\rf \Vert_1 + \lambda_{\rm{u}} \Vert \tilde{u} \Vert_1 + \lambda_{\rm{\tau}} \Vert \tilde{\tau} \Vert_1 , \label{eq:smpcj} \\
		& \text{s.t.} && s(k+ i | k) = \notag \\
		& && \hspace{0.02\textwidth}[f\e]\mmps (s(k + i -1 | k),u(k + i -1)), \label{eq:smpcc1}\\
		& && \hspace{0.21\textwidth} \forall i \in \{1, \dots, \Np\}, \notag\\
		& && [g]\mmps \left(s(k+ i -1 | k),u(k+i - 1)\right) \leqslant 0, \label{eq:smpcc4}\\
		& && \hspace{0.21\textwidth} \forall i \in \{1, \dots, \Np\}, \notag\\		
		& &&  \tau(i) = \min_{j = 1 \dots N_{\rm{lane}}} \left\{ \vert y\e (k + i | k) - y_{\rm{c}_j}\vert \right\} ,\label{eq:smpcc2}\\ 
		& && \hspace{0.21\textwidth} \forall i \in \{1, \dots, \Np\}, \notag\\
		& && [\mathbb{P}]\mmps^{(\eta)} (s(k + i | k)) \leqslant \epsilon, \label{eq:smpcc3}\\
		& && \hspace{0.21\textwidth} \forall \eta \in \{1, \dots, \No\}, \notag\\
		& && \hspace{0.21\textwidth} \forall i \in \{1, \dots, \Np\}. \notag
	\end{align}
\end{subequations}
where $[f\e]\mmps$ represents the discretized form of the MMPS system dynamics in (\ref{eq:evmmps}) and similarly, $[g]\mmps$ approximates the nonlinear constraints such as the Kamm circle. The objective is to minimize the cost in (\ref{eq:smpcj}) which consists of the collision probability, the deviation from the reference velocity, and the control effort. Moreover, the lane-center deviation $\tau$ is defined over the prediction horizon as (\ref{eq:smpcc2}) which allows switching to a ``better" lane (among $N_{\rm{lane}}$ lanes) if necessary. Here, $y_{\rm{c}_j}$ values represent the center line in lanes 1 and 2 for as two available lanes for the vehicle on the road and can be easily extended to include more lanes.  Constraints (\ref{eq:smpcc1}) and (\ref{eq:smpcc3}) respectively account for the prediction model of the ego vehicle and the chance constraints. The Proactive SMPC (P--SMPC) problem is solved via Algorithms~\ref{alg:prob} and~\ref{alg:smpc}.
\begin{algorithm}
	\caption{Probability function development}\label{alg:prob}
	\begin{algorithmic}	
		\Require $Z(k), \Xi_0, \Np$	\Comment{$Z$ \emph{contains states of all the obstacles}}
		\For{$\eta \in \{1, \dots, \No\}$}	
		\State $z^{(\eta)} (k) \gets \eta^{\rm{th}}$ column in $Z(k)$
		\State $\Xi^{(\eta)} (k | k) \gets \Xi_0$
		\For{$i \in \{1, \dots, \Np\}$}  \Comment{\emph{obstacle prediction}}
		\State $\Xi^{(\eta)} (k+i | k) \xleftarrow{\text{ (\ref{eq:obscov}) }} \Xi_0, \Xi^{(\eta)} (k + i -1 | k)$
		\State $\xi^{(\eta)} (k+i | k)  \xleftarrow{\text{ (\ref{eq:obsmodel}) }} z^{(\eta)} (k + i -1 | k)$ 
		\State $p_i^{(\eta)} \xleftarrow{\text{ (\ref{eq:pdfbeta}) }} \xi^{(\eta)} (k+i | k) ,\Xi^{(\eta)} (k+i | k)$ 
		\State $\mathbb{P}^{(\eta)} (.) \xleftarrow{\text{ develop using (\ref{eq:probeta}) }} p_i^{(\eta)}$
		\State $[\mathbb{P}]^{(\eta)}\mmps (.) \xleftarrow{\text{ (\ref{eq:parmp}) }} \mathbb{P}^{(\eta)} (.)$
		\State $[\hat{\mathbb{P}}]^{(\eta)}\mmps (.) \xleftarrow{\text{ (\ref{eq:parproxy}) }} \mathbb{P}^{(\eta)} (.)$
		\EndFor
		\EndFor
		\State \Return $[\mathbb{P}]^{(\eta)}\mmps (.), [\hat{\mathbb{P}}]^{(\eta)}\mmps (.) \quad \forall \eta \in \{1, \dots, \No\}$ 
	\end{algorithmic}
\end{algorithm}
\begin{algorithm}
	\caption{P--SMPC planner}\label{alg:smpc}
	\begin{algorithmic}	
		\Require $s(k), [f\e]\mmps, Z(k), \Xi_0, \tilde{v}\rf, \Np, y_{\rm{c}}$	
		\State $N_{\rm{lane}} \gets $ length of $y_{\rm{c}}$
		\State $[\mathbb{P}]\mmps (.), [\hat{\mathbb{P}}]\mmps (.) \xleftarrow{\text{ Algorithm~\ref{alg:prob} }} Z(k), \Xi_0, \Np$
		\State $P (.) \xleftarrow{\text{ develop using (\ref{eq:probfull}) }} [\hat{\mathbb{P}}]\mmps (.), \Np$
		\State $\tilde{s}^\ast \gets$ solve (\ref{eq:smpc}) \Comment{\emph{the planning optimization problem}}
		\State \Return $\tilde{s}^\ast$ 
	\end{algorithmic}
\end{algorithm}
\begin{rem}
	The chance constraints in the SMPC literature~\cite{Cannon2009} are often expressed by the generic form in (\ref{eq:probc2}). In our planner formulation, we use (\ref{eq:smpcc3}) as a more tractable formulation of chance constraints, which is essentially equivalent to bounding the constraint violation probability in (\ref{eq:probeta}) or its MMPS approximation (\ref{eq:parmp}) by a small value $\epsilon$. Note that $[\mathbb{P}]\mmps$ over-estimates $\mathbb{P}$ for probabilities close to zero as shown in {Fig.~\ref{fig:prob}}, and that in {(\ref{eq:smpcc3})} we make sure the collision probability is smaller than $\epsilon$ for all the states in $\mathcal{D}$ and all the time steps within the prediction horizon.
\end{rem}


\section{Simulations and Results}\label{sec:sim}

In this section, we evaluate the control performance of our proposed P--SMPC planner on two aspects: proactivity of the planner, and attainable generated trajectories. Here, we select $\epsilon = 0.001$ which is the tightest bound investigated in~\cite{Brudigam2023}. The P--SMPC optimization problem defines 10 continuous and 20 binary decision variables per prediction step to model the ego vehicle. Further, each detected obstacle adds up to 6 binary variables per prediction step to allow for hybrid representation of the collision probability function associated with it.

The proactivity assessment is done in four highway scenarios where we investigate the effect of collision-risk minimization in the objective function (\ref{eq:smpcj}) in our P--SMPC planner against the optimization formulation inspired by the state-of-the-art~\cite{Brudigam2023} indicated as \mbox{Regular SMPC (R--SMPC)} planner where the collision-risk is not included in the objective function and the collision is avoided by only considering the left-hand side of~(\ref{eq:smpcc3}). Note that \mbox{R--SMPC} is not the same planner as in~\cite{Brudigam2023} since it incorporates the MMPS approximation of the nonlinearities, but} we only change the objective function while keeping the same dynamic prediction model for both planners for a fair comparison and a better analysis of the risk-minimization effects. Further, we simulate the SMPC optimization problem in its nonlinear form as \mbox{Nonlinear SMPC (N--SMPC)} to compare the computation time against its MILP counterpart, P--SMPC. However, N--SMPC becomes infeasible in the complex scenario, which is discussed in more detail later.

To assess if the generated trajectory can be attained, we provide the reference trajectories provided by the P--SMPC planner to a high-fidelity vehicle model in IPG CarMaker~\cite{IPG} and compare the position and velocity trajectories of the ego vehicle with their references. 

The control frequency for all the simulations is set to 1kHz in accordance with the real-life applications where the computational capabilities limit the operational frequency of (digital) controllers~\cite{Stano2023}. The SMPC problems are all designed with sampling time of 0.2s and $\Np = 10$. We solve the MILPs using the GUROBI~\cite{GUROBI} optimizer and the NLPs using the SQP solver in \texttt{fmincon} in a \textsc{Matlab} R2020b environment. For a fair comparison between the two solvers, we provide the objective and the constraints as object code to speed-up the solution time of the NLPs, which in our simulations, has resulted in up to 20 times faster convergence compared to providing the objective and the constraints as \textsc{Matlab} functions. The simulations were run on a PC with a 8-core(s) Intel Xeon 3.60 GHz CPU and 8 GB RAM on Windows 10 64-bit. The codes are available from~\cite{SMPCcode}.

\subsection{Proactivity Assessment}

In real life, some of the most dangerous situations on a highway are sudden appearance of a static object or extreme deceleration of a front vehicle. Therefore, we define different conceptual scenarios with slow-moving vehicles in all of them to present scenarios where the obstacle is so slow (or even static) that slowing down to keep the distance for collision avoidance is either impossible for the ego vehicle or extremely dangerous. As a result, we can test the ability of the planner in finding a safe, yet aggressive, evasive maneuver to avoid the collision.  For this, the initial longitudinal velocity of the ego vehicle is considered to be $22$m/s ($\approx80$km/h), while the dynamic obstacles are assumed to have initial velocities between $8$ to $11$m/s ($\approx 30$ to $40$km/h). Nevertheless, we select the scenarios in a way to represent challenging, yet possible cases where e.g.\ other drivers do not aim to collide with each other, but may behave carelessly. 

We use four conceptual scenarios to assess the solutions of the P--SMPC planner:
\begin{enumerate}
	\item \textbf{Single obstacle: } A slow-moving obstacle is in front of the ego vehicle on the same lane. We expect the ego vehicle to avoid collision with this obstacle by performing an evasive maneuver, instead of slowing down to keep a safe distance.
	\item \textbf{Dynamic corridor:}  In addition to an obstacle in the lane as the single-obstacle scenario, there is another slow-moving vehicle on the other lane to present a situation where the ego vehicle needs to pass through a corridor between two dynamic obstacles with stochastic behavior. Here we expect the ego vehicle to pass that corridor along an optimal trajectory.
	\item \textbf{Static/dynamic corridor:} This scenario is similar to the dynamic corridor, except here we have a static object on the road instead of another slow-moving obstacle. 
	\item \textbf{Complex scenario:} Here we assess the planner in a situation where there are four slow-moving vehicles (two on each lane) and one static object present on the road. There exists a safe corridor between the dynamic and static obstacles, in which we expect the planner to find an optimal trajectory.
\end{enumerate}
Moreover, each scenario is investigated twice: first as realization (i) where the obstacles behave ideally as the P--SMPC planner calculates $z\rf$, i.e.\ they keep their longitudinal velocity and lateral position, and secondly as realization (ii) where some/all of them either change their speed or their lane. Note that in realization (ii), the obstacle's intention to change lanes is not known a priori to the ego vehicle, as a result, the SMPC planners keep the assumption that the obstacle behaves as realization (i).
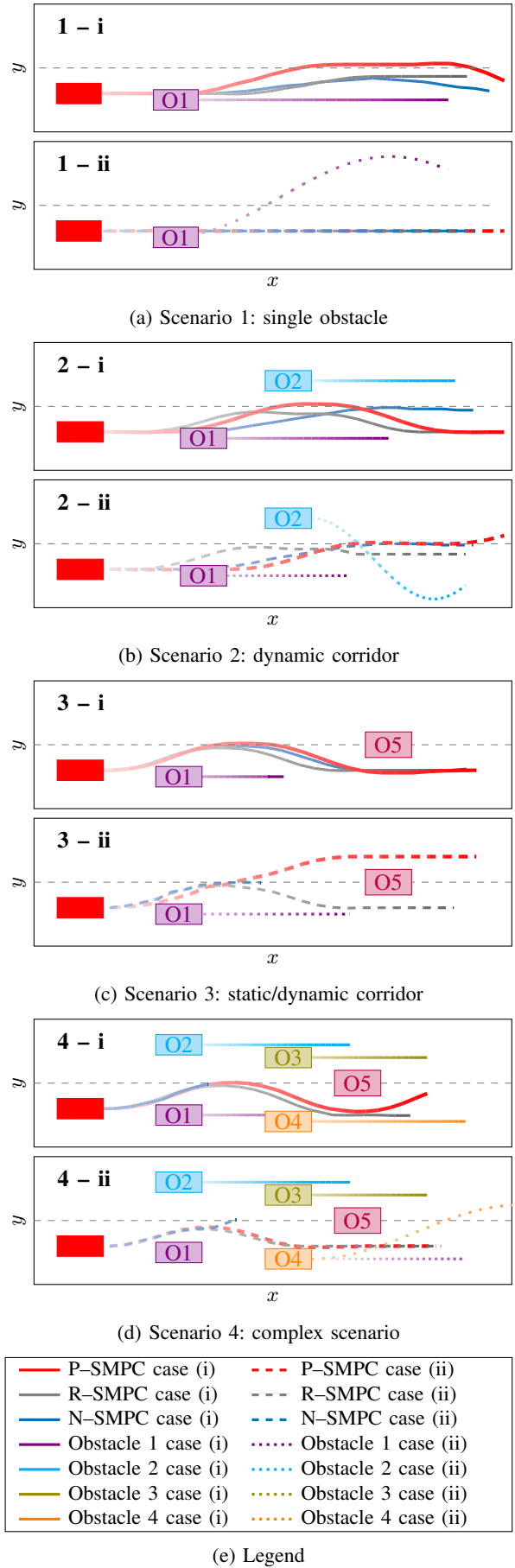
\begin{figure}[hbtp]
\begin{center}
\begin{subfigure}[b]{\linewidth}\centering
\begin{tikzpicture}
	\begin{axis}[width=\textwidth,height=0.4\textwidth,xtick=\empty, ytick=\empty, 
		xmin=-1.5,xmax=30.5,ymin=-5,ymax=5,xlabel=\empty, ylabel=$y$,]			
		\addplot[smooth,mesh,colormap={violetwhite}{color=(violet!10) color=(violet)},scatter src=\thisrow{T},very thick] table [x=ZX11,y=ZY11]{Data/PT_Scenario1.dat};
		\addplot[smooth,mesh,colormap={bluewhite}{color=(NavyBlue!10) color=(NavyBlue)},scatter src=\thisrow{T},very thick] table [x=XN1,y=YN1]{Data/PT_Scenario1.dat};
		\addplot[smooth,mesh,colormap={graywhite}{color=(gray!10) color=(black!60)},scatter src=\thisrow{T},very thick] table [x=XR1,y=YR1]{Data/PT_Scenario1.dat};
		\addplot[smooth,mesh,colormap={redwhite}{color=(red!10) color=(red)},scatter src=\thisrow{T},ultra thick]	table [x=XP1,y=YP1]{Data/PT_Scenario1.dat};
		\draw[red,fill=red,thin] (0,-2.8) rectangle (3,-1.2);
		\draw[gray,dashed] (-3,0)--(29,0);
		\draw[violet,fill=violet!30,thin] (6.5,-3.3) rectangle (9.5,-1.7) node[midway] {O1};	
		\node[right] at (-0.5,3.3) {\textbf{1 -- i}};
	\end{axis}
\end{tikzpicture}\vspace{-0.1cm}
\begin{tikzpicture}
	\begin{axis}[width=\textwidth,height=0.4\textwidth,xtick=\empty, ytick=\empty, 
		xmin=-1.5,xmax=30.5,ymin=-5,ymax=5,xlabel=$x$, ylabel=$y$,]			
		\addplot[smooth,mesh,colormap={redwhite}{color=(red!10) color=(red)},scatter src=\thisrow{T},ultra thick,dashed] table [x=XP2,y=YP2]{Data/PT_Scenario1.dat};
		\addplot[smooth,mesh,colormap={graywhite}{color=(gray!10) color=(black!60)},scatter src=\thisrow{T},very thick,dashed] table [x=XR2,y=YR2]{Data/PT_Scenario1.dat};
		\addplot[smooth,mesh,colormap={bluewhite}{color=(NavyBlue!10) color=(NavyBlue)},scatter src=\thisrow{T},very thick,dashed] table [x=XN2,y=YN2]{Data/PT_Scenario1.dat};
		\addplot[smooth,mesh,colormap={violetwhite}{color=(violet!20) color=(violet!99)},scatter src=\thisrow{T},very thick,loosely dotted] table [x=ZX12,y=ZY12]{Data/PT_Scenario1.dat};
		\draw[violet,fill=violet!30,thin] (6.5,-3.3) rectangle (9.5,-1.7) node[midway] {O1};	
		\draw[red,fill=red,thin] (0,-2.8) rectangle (3,-1.2);
		\draw[gray,dashed] (-3,0)--(29,0);
		\node[right] at (-0.5,3.3) {\textbf{1 -- ii}};
	\end{axis}
\end{tikzpicture}
\subcaption{Scenario 1: single obstacle}\label{fig:scen1}\vspace{5pt}
\end{subfigure}			
\begin{subfigure}[b]{\linewidth}\centering
\begin{tikzpicture}
	\begin{axis}[width=\textwidth,height=0.4\textwidth,xtick=\empty, ytick=\empty, 
		xmin=-3,xmax=28,ymin=-5,ymax=5,xlabel=\empty, ylabel=$y$,]			
		\addplot[smooth,mesh,colormap={violetwhite}{color=(violet!10) color=(violet)},scatter src=\thisrow{T},very thick] table [x=ZX11,y=ZY11]{Data/PT_Scenario2.dat};
		\addplot[smooth,mesh,colormap={cyanwhite}{color=(cyan!10) color=(cyan)},scatter src=\thisrow{T},very thick] table [x=ZX21,y=ZY21]{Data/PT_Scenario2.dat};
		\addplot[smooth,mesh,colormap={bluewhite}{color=(NavyBlue!10) color=(NavyBlue)},scatter src=\thisrow{T},very thick] table [x=XN1,y=YN1]{Data/PT_Scenario2.dat};
		\addplot[smooth,mesh,colormap={graywhite}{color=(gray!10) color=(black!60)},scatter src=\thisrow{T},very thick] table [x=XR1,y=YR1]{Data/PT_Scenario2.dat};
		\addplot[smooth,mesh,colormap={redwhite}{color=(red!10) color=(red)},scatter src=\thisrow{T},ultra thick]	table [x=XP1,y=YP1]{Data/PT_Scenario2.dat};
		\draw[red,fill=red,thin] (-1.5,-2.8) rectangle (1.5,-1.2);
		\draw[violet,fill=violet!30,thin] (6.5,-3.3) rectangle (9.5,-1.7) node[midway] {O1};
		\draw[gray,dashed] (-3,0)--(28,0);
		\draw[cyan,fill=cyan!30,thin] (12,1.2) rectangle (15,2.8) node[midway] {O2};
		\node[right] at (-2.0,3.3) {\textbf{2 -- i}};
	\end{axis}
\end{tikzpicture}\vspace{-0.1cm}
\begin{tikzpicture}
	\begin{axis}[width=\textwidth,height=0.4\textwidth,xtick=\empty, ytick=\empty, 
		xmin=-3,xmax=28,ymin=-5,ymax=5,xlabel=$x$, ylabel=$y$,]			
		\addplot[smooth,mesh,colormap={bluewhite}{color=(NavyBlue!10) color=(NavyBlue)},scatter src=\thisrow{T},very thick,dashed] table [x=XN2,y=YN2]{Data/PT_Scenario2.dat};
		\addplot[smooth,mesh,colormap={redwhite}{color=(red!10) color=(red)},scatter src=\thisrow{T},ultra thick,dashed] 
		table [x=XP2,y=YP2]{Data/PT_Scenario2.dat};
		\addplot[smooth,mesh,colormap={cyanwhite}{color=(cyan!10) color=(cyan)},scatter src=\thisrow{T},very thick,dotted] table [x=ZX22,y=ZY22]{Data/PT_Scenario2.dat};
		\addplot[smooth,mesh,colormap={graywhite}{color=(gray!10) color=(black!60)},scatter src=\thisrow{T},very thick,dashed] table [x=XR2,y=YR2]{Data/PT_Scenario2.dat};
		\addplot[smooth,mesh,colormap={violetwhite}{color=(violet!10) color=(violet)},scatter src=\thisrow{T},very thick,dotted] table [x=ZX12,y=ZY12]{Data/PT_Scenario2.dat};
		\draw[red,fill=red,thin] (-1.5,-2.8) rectangle (1.5,-1.2);
		\draw[violet,fill=violet!30,thin] (6.5,-3.3) rectangle (9.5,-1.7) node[midway] {O1};
		\draw[gray,dashed] (-3,0)--(28,0);
		\draw[cyan,fill=cyan!30,thin] (12,1.2) rectangle (15,2.8) node[midway] {O2};
		\node[right] at (-2.0,3.3) {\textbf{2 -- ii}};
	\end{axis}
\end{tikzpicture}
\subcaption{Scenario 2: dynamic corridor}\label{fig:scen2}\vspace{5pt}
\end{subfigure}			
\begin{subfigure}[b]{\linewidth}\centering
\begin{tikzpicture}
	\begin{axis}[width=\textwidth,height=0.4\textwidth,xtick=\empty, ytick=\empty, 
		xmin=-3,xmax=28,ymin=-5,ymax=5,xlabel=\empty, ylabel=$y$,]	
		\addplot[smooth,mesh,colormap={bluewhite}{color=(NavyBlue!10) color=(NavyBlue)},scatter src=\thisrow{T},very thick] table [x=XN1,y=YN1]{Data/PT_Scenario3.dat};		
		\addplot[smooth,mesh,colormap={violetwhite}{color=(violet!10) color=(violet)},scatter src=\thisrow{T},very thick] table [x=ZX11,y=ZY11]{Data/PT_Scenario3.dat};
		\addplot[smooth,mesh,colormap={graywhite}{color=(gray!10) color=(black!60)},scatter src=\thisrow{T},very thick] table [x=XR1,y=YR1]{Data/PT_Scenario3.dat};
		\addplot[smooth,mesh,colormap={redwhite}{color=(red!10) color=(red)},scatter src=\thisrow{T},ultra thick] 
		table [x=XP1,y=YP1]{Data/PT_Scenario3.dat};
		\draw[red,fill=red,thin] (-1.5,-2.8) rectangle (1.5,-1.2);
		\draw[violet,fill=violet!30,thin] (4.9,-3.3) rectangle (7.9,-1.7) node[midway] {O1};			
		\draw[gray,dashed] (-3,0)--(28,0);
		\draw[purple,fill=purple!30,thin] (18.5,-1) rectangle (21.5,1) node[midway] {O5};
		\node[right] at (-2.0,3.3) {\textbf{3 -- i}};
	\end{axis}
\end{tikzpicture}\vspace{-0.1cm}
\begin{tikzpicture}
	\begin{axis}[width=\textwidth,height=0.4\textwidth,xtick=\empty, ytick=\empty, 
		xmin=-3,xmax=28,ymin=-5,ymax=5,xlabel=$x$, ylabel=$y$,]			
		\addplot[smooth,mesh,colormap={redwhite}{color=(red!10) color=(red)},scatter src=\thisrow{T},ultra thick,dashed] table [x=XP2,y=YP2]{Data/PT_Scenario3.dat};
		\addplot[smooth,mesh,colormap={graywhite}{color=(gray!10) color=(black!60)},scatter src=\thisrow{T},very thick,dashed] table [x=XR2,y=YR2]{Data/PT_Scenario3.dat};
		\addplot[smooth,mesh,colormap={bluewhite}{color=(NavyBlue!30) color=(NavyBlue)},scatter src=\thisrow{T},very thick,dashed] table [x=XN2,y=YN2]{Data/PT_Scenario3.dat};		
		\addplot[smooth,mesh,colormap={violetwhite}{color=(violet!10) color=(violet)},scatter src=\thisrow{T},very thick,dotted] 
		table [x=ZX12,y=ZY12]{Data/PT_Scenario3.dat};
		\draw[red,fill=red,thin] (-1.5,-2.8) rectangle (1.5,-1.2);
		\draw[violet,fill=violet!30,thin] (4.9,-3.3) rectangle (7.9,-1.7) node[midway] {O1};			
		\draw[gray,dashed] (-3,0)--(28,0);
		\draw[purple,fill=purple!30,thin] (18.5,-1) rectangle (21.5,1) node[midway] {O5};
		\node[right] at (-2.0,3.3) {\textbf{3 -- ii}};
	\end{axis}
\end{tikzpicture}
\subcaption{Scenario 3: static/dynamic corridor}\label{fig:scen3}\vspace{5pt}
\end{subfigure}			
\begin{subfigure}[b]{\linewidth}\centering
\begin{tikzpicture}
	\begin{axis}[width=\textwidth,height=0.4\textwidth,xtick=\empty, ytick=\empty, 
		xmin=-3,xmax=28,ymin=-5,ymax=5,xlabel=\empty, ylabel=$y$,]			
		\draw[gray,dashed] (-3,0)--(28,0);
		\addplot[smooth,mesh,colormap={redwhite}{color=(red!10) color=(red)},scatter src=\thisrow{T},ultra thick] 
		table [x=XP1,y=YP1]{Data/PT_Scenario4.dat};
		\addplot[smooth,mesh,colormap={cyanwhite}{color=(cyan!10) color=(cyan)},scatter src=\thisrow{T},very thick] 
		table [x=ZX31,y=ZY31]{Data/PT_Scenario4.dat};
		\addplot[smooth,mesh,colormap={graywhite}{color=(gray!10) color=(black!60)},scatter src=\thisrow{T},very thick] 
		table [x=XR1,y=YR1]{Data/PT_Scenario4.dat};
		\addplot[smooth,mesh,colormap={violetwhite}{color=(violet!01) color=(violet!70)},scatter src=\thisrow{T},very thick] 
		table [x=ZX11,y=ZY11]{Data/PT_Scenario4.dat};
		\addplot[smooth,mesh,colormap={orangewhite}{color=(orange!30) color=(orange!90)},scatter src=\thisrow{T},very thick] 
		table [x=ZX41,y=ZY41]{Data/PT_Scenario4.dat};		
		\addplot[smooth,mesh,colormap={olivewhite}{color=(olive!10) color=(olive)},scatter src=\thisrow{T},very thick] 
		table [x=ZX51,y=ZY51]{Data/PT_Scenario4.dat};		
		\addplot[smooth,mesh,colormap={bluewhite}{color=(NavyBlue!30) color=(NavyBlue)},scatter src=\thisrow{T},very thick] table [x=XN1,y=YN1]{Data/PT_Scenario4.dat};		
		\draw[red,fill=red,thin] (-1.5,-2.8) rectangle (1.5,-1.2);
		\draw[violet,fill=violet!30,thin] (4.9,-3.3) rectangle (7.9,-1.7) node[midway] {O1};
		\draw[purple,fill=purple!30,thin] (16.5,-1) rectangle (19.5,1) node[midway] {O5};
		\draw[cyan,fill=cyan!30,thin] (4.9,2.2) rectangle (7.9,3.8) node[midway] {O2};
		\draw[orange,fill=orange!30,thin] (12,-3.8) rectangle (15,-2.2) node[midway] {O4};
		\draw[olive,fill=olive!30,thin] (12,1.2) rectangle (15,2.8) node[midway] {O3};		
		\node[right] at (-2.0,3.3) {\textbf{4 -- i}};
	\end{axis}
\end{tikzpicture}\vspace{-0.1cm}
\begin{tikzpicture}
	\begin{axis}[width=\textwidth,height=0.4\textwidth,xtick=\empty, ytick=\empty, 
		xmin=-3,xmax=28,ymin=-5,ymax=5,xlabel=$x$, ylabel=$y$,]			
		\draw[gray,dashed] (-3,0)--(28,0);
		\addplot[smooth,mesh,colormap={redwhite}{color=(red!10) color=(red)},scatter src=\thisrow{T},ultra thick,dashed] 
		table [x=XP2,y=YP2]{Data/PT_Scenario4.dat};
		\addplot[smooth,mesh,colormap={cyanwhite}{color=(cyan!10) color=(cyan)},scatter src=\thisrow{T},very thick,dashed] 
		table [x=ZX31,y=ZY31]{Data/PT_Scenario4.dat};
		\addplot[smooth,mesh,colormap={graywhite}{color=(gray!10) color=(black!60)},scatter src=\thisrow{T},very thick,dashed] 
		table [x=XR2,y=YR2]{Data/PT_Scenario4.dat};
		\addplot[smooth,mesh,colormap={violetwhite}{color=(violet!01) color=(violet!70)},scatter src=\thisrow{T},very thick,dotted] 
		table [x=ZX12,y=ZY12]{Data/PT_Scenario4.dat};
		\addplot[smooth,mesh,colormap={orangewhite}{color=(orange!40) color=(orange!90)},scatter src=\thisrow{T},very thick,loosely dotted] 
		table [x=ZX42,y=ZY42]{Data/PT_Scenario4.dat};		
		\addplot[smooth,mesh,colormap={olivewhite}{color=(olive!10) color=(olive)},scatter src=\thisrow{T},very thick,dashed] 
		table [x=ZX52,y=ZY52]{Data/PT_Scenario4.dat};		
		\addplot[smooth,mesh,colormap={bluewhite}{color=(NavyBlue!30) color=(NavyBlue)},scatter src=\thisrow{T},very thick,dashed] table [x=XN2,y=YN2]{Data/PT_Scenario4.dat};		
		\draw[red,fill=red,thin] (-1.5,-2.8) rectangle (1.5,-1.2);
		\draw[violet,fill=violet!30,thin] (4.9,-3.3) rectangle (7.9,-1.7) node[midway] {O1};
		\draw[purple,fill=purple!30,thin] (16.5,-1) rectangle (19.5,1) node[midway] {O5};
		\draw[cyan,fill=cyan!30,thin] (4.9,2.2) rectangle (7.9,3.8) node[midway] {O2};
		\draw[orange,fill=orange!30,thin] (12,-3.8) rectangle (15,-2.2) node[midway] {O4};
		\draw[olive,fill=olive!30,thin] (12,1.2) rectangle (15,2.8) node[midway] {O3};		
		\node[right] at (-2.0,3.3) {\textbf{4 -- ii}};
	\end{axis}
\end{tikzpicture}
\subcaption{Scenario 4: complex scenario}\label{fig:scen4}\vspace{5pt}
\end{subfigure}			
\begin{subfigure}[b]{\linewidth}\centering
\begin{tikzpicture}
	\begin{axis}[width=1.05\textwidth,height=0.48\textwidth,xtick=\empty, ytick=\empty, 
		xmin=-1,xmax=36,ymin=-9,ymax=5,]			
		\draw[red,very thick] (0,4)--(3,4) node[right,midway,xshift=3mm,black] {\small P--SMPC case (i)};
		\draw[red,very thick,dashed] (17,4)--(20,4) node[right,midway,xshift=3mm,black] {\small P--SMPC case (ii)};
		\draw[gray,very thick] (0,2)--(3,2) node[right,midway,xshift=3mm,black] {\small R--SMPC case (i)};
		\draw[gray,very thick,dashed] (17,2)--(20,2) node[right,midway,xshift=3mm,black] {\small R--SMPC case (ii)};	
		\draw[NavyBlue,very thick] (0,0)--(3,0) node[right,midway,xshift=3mm,black] {\small N--SMPC case (i)};
		\draw[NavyBlue,very thick,dashed] (17,0)--(20,0) node[right,midway,xshift=3mm,black] {\small N--SMPC case (ii)};				
		\draw[violet,very thick] (0,-2)--(3,-2) node[right,midway,xshift=3mm,black] {\small Obstacle 1 case (i)};
		\draw[violet,very thick,dotted] (17,-2)--(20,-2) node[right,midway,xshift=3mm,black] {\small Obstacle 1 case (ii)};
		\draw[cyan,very thick] (0,-4)--(3,-4) node[right,midway,xshift=3mm,black] {\small Obstacle 2 case (i)};
		\draw[cyan,very thick,dotted] (17,-4)--(20,-4) node[right,midway,xshift=3mm,black] {\small Obstacle 2 case (ii)};
		\draw[olive,very thick] (0,-6)--(3,-6) node[right,midway,xshift=3mm,black] {\small Obstacle 3 case (i)};
		\draw[olive,very thick,dotted] (17,-6)--(20,-6) node[right,midway,xshift=3mm,black] {\small Obstacle 3 case (ii)};
		\draw[orange,very thick] (0,-8)--(3,-8) node[right,midway,xshift=3mm,black] {\small Obstacle 4 case (i)};
		\draw[orange,very thick,dotted] (17,-8)--(20,-8) node[right,midway,xshift=3mm,black] {\small Obstacle 4 case (ii)};
	\end{axis}		
\end{tikzpicture}
\subcaption{Legend}\label{fig:scenleg}
\end{subfigure}			
\caption{Simulation results for proactivity assessment of the planners. The ego vehicle is shown by a red rectangle and the fading represents the trajectory evolution over time. Note that obstacle 5 (O5) is static.}\label{fig:scenrarios}
\end{center}
\end{figure}

In total, we have conducted 400 Monte-Carlo simulations by perturbing the initial speed and the longitudinal distance between the ego vehicle and the obstacles with uniform sampling within a $\pm 5$\% range as an acceptable bound from the literature~\cite{Chai2020}. In Fig.~\ref{fig:scenrarios}, four examples are selected as most clear cases to showcase the efficacy of our approach in a more clear way. The statistical information regarding the Monte-Carlo simulations can be found in Fig.~\ref{fig:costc}. The ego vehicle is shown in red, while the obstacles are labeled by the letter ``O" and a number to distinguish among them. The solid lines represent the case where the obstacles move according to the obstacle prediction model and keep their longitudinal velocity and lateral position. The dashed lines correspond to a case where the obstacles behave differently than the obstacle prediction model in P--SMPC, e.g.\ some of the obstacles on the road are accelerating/decelerating or intending to change their lanes. The solid red line shows the generated reference trajectory in cases with realization (i), while the dashed one shows the solution in realization (ii). R--SMPC and N--SMPC results are shown respectively in gray and blue in a similar fashion. 

\subsubsection{Single obstacle}

In the first scenario (Fig.~\ref{fig:scen1}), the P--SMPC and R--SMPC planners avoid collision when the obstacle behaves as predicted by an evasive maneuver. However, P--SMPC planner keeps a larger distance with a higher speed compared to the R--SMPC planner that converges to a trajectory that only satisfies the chance constraints (left-hand side of (\ref{eq:smpcc3})) and favors a solution that is closer to the middle of the lane. Note that the higher average velocity is visible by comparing the length of the red and gray trajectories. If the obstacle intends to change lanes which is not known to the ego vehicle a priori, the P--SMPC and R--SMPC planners both keep on assuming that the obstacle will keep its lateral position in each control step, but after the initial control steps, the planners converge to trajectories on the same lane as the obstacle merges into the next one. The difference between the planners is that the P--SMPC converges to slightly higher speed (since the red dashed line extends more to the right) to keep more distance from the obstacle. In this sense, P--SMPC is more proactive as its manages to get out of the hazardous situation while ensuring a higher safety level.

\subsubsection{Dynamic corridor}

Figure~\ref{fig:scen2} shows the simulation results during the dynamic corridor scenario where both obstacles are moving. If the obstacles behave as predicted by the ego vehicle and intend to keep driving on the same lane, the P--SMPC and R--SMPC planners avoid the collision by overtaking O1 and returning to the center of the right lane. Here, the P--SMPC planner keeps more distance with O1 since it succeeds in finding a trajectory that has a lower collision risk than the left-hand side of (\ref{eq:smpcc3}). however, if O2 actually intends to move to the right lane, after a few control steps when the ego vehicle observes the updated lateral position of O2, P--SMPC keeps more distance from the center of the right lane and eventually merges into the left lane as it detects this area to be the safest option. It should be noted that this is possible due to allowing switching between lanes in (\ref{eq:smpcc2}). Otherwise, the planners would keep aiming for staying on the right lane which means driving on the center line between the two lanes until the right lane is risk-free. The R--SMPC, however, is not able to use this potential since it keeps a closer trajectory to the obstacles and does not search for other trajectories with lower collision-risk, as long as (\ref{eq:smpcc3}) is satisfied. As a result, P--SMPC is more proactive in the sense of avoiding the propagation of hazard to the next time steps.

\subsubsection{Static/dynamic corridor}

In the dynamic/static corridor scenario, both P--SMPC and R--SMPC planners avoid colliding with the obstacles by overtaking O1 as shown in Fig.~\ref{fig:scen3}, where P--SMPC planner keeps a larger distance with the ``more uncertain" obstacle (O1). However, if O1 intends to increase its longitudinal velocity, R--SMPC planner still converges to the same trajectory since it still satisfies the (\ref{eq:smpcc3}), whereas P--SMPC changes lanes to the safer track and avoids the collision by overtaking the static obstacle O5 from the left. Similar to the dynamic corridor, this may lead to hazard propagation to the next steps, a problem which P--SMPC mitigates by proactive collision avoidance via finding a solution with a lower collision risk for future time steps.

\subsubsection{Complex scenario}

Figure~\ref{fig:scen4} shows the simulations for the complex scenario. If obstacles behave as predicted by the ego vehicle, the P--SMPC and R--SMPC planners manage to find a solution within the attainable corridor to avoid collision with the road users. In the final control steps, the left lane is empty and safer, therefore P--SMPC planner decides to merge to the left lane, whereas R--SMPC keeps the same lane. However, if O1 steers to the right and O4 intends to merge into the left lane, P--SMPC planner decides to stay in the same lane as the right lane is the safer one and suggests a similar trajectory as planned by the R--SMPC planner. Figure{~\ref{fig:appx} shows the force plots during the complex scenario as an example to show the capability of the SMPC to operate close to the tire saturation limits. Note that the velocity of the ego vehicle during the maneuvers is not always constant and is discussed in more detail in the next section, accompanied by corresponding plots.
\begin{figure}[hbtp]
\begin{center}
\begin{tikzpicture}
	\begin{axis}[width=0.3\textwidth,height=0.3\textwidth,
		xmajorgrids=true,xminorgrids=true,ymajorgrids=true,yminorgrids=true,grid style=dashed,xtick={-3000,0,3000},ytick={-3000,0,3000},
		xmin=-7500,xmax=7500,ymin=-7500,ymax=7500,legend pos=outer north east,
		xlabel=$F_x$ (N), ylabel=$F_y$ (N),]			
		\addplot[smooth,color=Maroon,very thick] table [x=FXF,y=FYF]{Data/Appendix.dat};
		\addplot[smooth,color=TealBlue,very thick] table [x=FXR,y=FYR]{Data/Appendix.dat};
		\draw[dashed,gray] (0,0) circle [radius=7200];
		\legend{Front,Rear};
	\end{axis}
\end{tikzpicture}
\caption{Force plot of the complex maneuver ({Fig.~\ref{fig:scen4}}) with the Kamm circle shown by dashed line.}\label{fig:appx}
\end{center}
\end{figure}
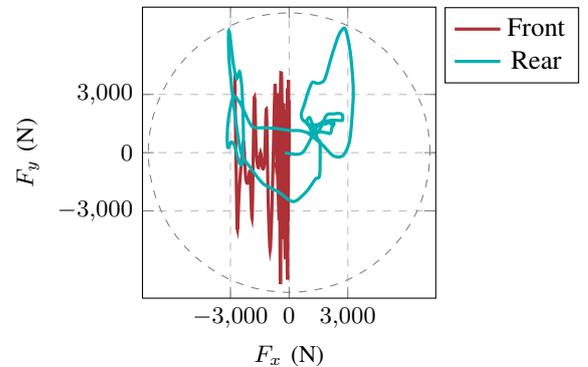
Note that the N--SMPC planner reaches infeasibility before the end of simulations in the last three cases, which leads to incomplete trajectories. This phenomenon is a result of using a warm start strategy (or solution using limited and insufficient number of initial guesses) which in turn leads to accumulation of errors after a few time steps as follows: in the complex scenario, the ego vehicle detects the obstacles 2 seconds before reaching their current position, e.g. O4 is detected after the ego initiates steering to avoid colliding with O1. Using the shifted solution of the previous time step in such cases leads to a poor result: as the previous solution was to go back to the initial lateral position after overtaking O1, by detecting O4, the planner converges to a solution that suggest going back to the initial lateral position after overtaking O4. Conversely, R–SMPC and P–SMPC planners are able to find a better solution thanks to their search for a global optimum, which is to brake and steer to the center of the lane to keep more distance from O4. In the next time steps, O5 is detected, and R--SMPC and P--SMPC manage to find a trajectory to steer to the center of the lane faster now that an obstacle is in the way. However, the poor solution in the previous time steps from the N–SMPC planner has resulted in higher longitudinal velocity. Therefore, the time to collision with O5 is shorter and it is infeasible to find a trajectory to avoid colliding with O5 with the current velocity.
\begin{rem}
	In some cases, the N--SMPC may not be able to converge to a solution before the next time step, which means the best feasible point found by the solver during iterations will be used.
\end{rem}

\subsection{Assessment of Attainable Trajectories}\label{sec:attain}

To assess if the trajectories generated by the planner can be attained, we check whether they can be tracked by a high-fidelity vehicle model. In the proactivity test, eight reference trajectories were generated in total by the P--SMPC planner. To avoid repetition, we select four of these trajectories as distinct maneuvers and we simulate the high-fidelity BMW model in IPG CarMaker~\cite{IPG} to track them. It should be noted that the other trajectories produced similar results. The selected maneuvers are:
\begin{enumerate}[label=\alph*)]
	\item \textbf{Constant-speed overtake:} scenario (2-i), the solid red line in Fig.~\ref{fig:scen2},
	\item \textbf{Decelerating overtake:} scenario (3-i), the solid red line in Fig.~\ref{fig:scen3},
	\item \textbf{Double overtake:} scenario (3-ii), the dashed red line in Fig.~\ref{fig:scen3}, and
	\item \textbf{Lane change:} scenario (2-ii), the dashed red line in Fig.~\ref{fig:scen2}.
\end{enumerate}
In each simulation, we give the velocity vector in the four maneuvers to the longitudinal controller in IPG as the reference velocity profile, and provide the steering angles to the lateral controller for lateral motion. Figure~\ref{fig:ipgsim} shows comparisons of the $x\e$, $y\e$, and $v$ trajectories obtained by the P--SMPC planner and the resulting trajectory of the IPG vehicle.
\begin{rem}
	We start each IPG simulation from $x\e=0m$ and run a steady, constant velocity maneuver for 200m to allow for the IPG model to stabilize before tracking the reference maneuver. As a result, the attainability tests start at \mbox{$x\e =200$m}. 
\end{rem}

Figure~\ref{fig:ipgsim} shows that the reference trajectories provided by P--SMPC planner are attainable for the high-fidelity IPG model to track, with slight mismatch along the $X$ axis, which is reasonable considering the larger complexity of the higher-fidelity model in IPG CarMaker, as compared to the prediction model in the P--SMPC planner.
\begin{figure}[hbtp]
\begin{center}
\begin{subfigure}[b]{\linewidth}\centering
\begin{tikzpicture}
	\begin{axis}[width=0.62\textwidth,height=0.4\textwidth,ytick={-5,0,5},legend columns=2,
		legend style={draw=none,fill=none,at={(0.55,0.9)},anchor=north},
		xmin=220,xmax=350,ymin=-6,ymax=6,xlabel=$x\e$ (m), ylabel=$y\e$ (m),]	
		\draw[gray,dashed] (200,0)--(400,0);		
		\draw[gray] (200,5)--(400,5);		
		\draw[gray] (200,-5)--(400,-5);		
		\addplot[smooth,color=Fuchsia,very thick] table [x=X,y=Y]{Data/CM_OTCV_REF.dat};
		\addplot[smooth,color=Dandelion,very thick] table [x=X,y=Y]{Data/CM_OTCV_IPG.dat};
		\legend{Planner, IPG}
	\end{axis}
\end{tikzpicture}
\begin{tikzpicture}
	\begin{axis}[width=0.38\textwidth,height=0.4\textwidth,
		xmajorgrids=true,xminorgrids=true,ymajorgrids=true,yminorgrids=true,grid style=dashed,xtick={1,3,5},ytick={20,21,22},
		xmin=0,xmax=5,ymin=19,ymax=23,xlabel=$t$ (s), ylabel=$v$ (m/s),]			
		\addplot[smooth,color=Fuchsia,very thick] table [x=T,y=V]{Data/CM_OTCV_REF.dat};
		\addplot[smooth,color=Dandelion,very thick] table [x=T,y=V]{Data/CM_OTCV_IPG.dat};
	\end{axis}
\end{tikzpicture}
\subcaption{Constant-speed overtake}\label{fig:ipg1}\vspace{5pt}
\end{subfigure}			
\begin{subfigure}[b]{\linewidth}\centering
\begin{tikzpicture}
	\begin{axis}[width=0.62\textwidth,height=0.4\textwidth,ytick={-5,0,5},legend columns=2,
		legend style={draw=none,fill=none,at={(0.55,0.9)},anchor=north},
		xmin=220,xmax=350,ymin=-6,ymax=6,xlabel=$x\e$ (m), ylabel=$y\e$ (m),]	
		\draw[gray,dashed] (200,0)--(400,0);		
		\draw[gray] (200,5)--(400,5);		
		\draw[gray] (200,-5)--(400,-5);		
		\addplot[smooth,color=Fuchsia,very thick] table [x=X,y=Y]{Data/CM_OTVV_REF.dat};
		\addplot[smooth,color=Dandelion,very thick] table [x=X,y=Y]{Data/CM_OTVV_IPG.dat};
		\legend{Planner, IPG}
	\end{axis}
\end{tikzpicture}
\begin{tikzpicture}
	\begin{axis}[width=0.38\textwidth,height=0.4\textwidth,
		xmajorgrids=true, xminorgrids=true,ymajorgrids=true,yminorgrids=true,grid style=dashed,xtick={1,3,5},ytick={20,21,22},
		xmin=0,xmax=5,ymin=19,ymax=23,xlabel=$t$ (s), ylabel=$v$ (m/s),]					
		\addplot[smooth,color=Fuchsia,very thick] table [x=T,y=V]{Data/CM_OTVV_REF.dat};
		\addplot[smooth,color=Dandelion,very thick] table [x=T,y=V]{Data/CM_OTVV_IPG.dat};
	\end{axis}
\end{tikzpicture}
\subcaption{Decelerating overtake}\label{fig:ipg2}\vspace{5pt}
\end{subfigure}			
\begin{subfigure}[b]{\linewidth}\centering
\begin{tikzpicture}
	\begin{axis}[width=0.62\textwidth,height=0.4\textwidth,ytick={-5,0,5},legend columns=2,
		legend style={draw=none,fill=none,at={(0.55,0.37)},anchor=north},
		xmin=220,xmax=350,ymin=-6,ymax=6,xlabel=$x\e$ (m), ylabel=$y\e$ (m),]	
		\draw[gray,dashed] (200,0)--(400,0);		
		\draw[gray] (200,5)--(400,5);		
		\draw[gray] (200,-5)--(400,-5);		
		\addplot[smooth,color=Fuchsia,very thick] table [x=X,y=Y]{Data/CM_DOOT_REF.dat};
		\addplot[smooth,color=Dandelion,very thick] table [x=X,y=Y]{Data/CM_DOOT_IPG.dat};
		\legend{Planner, IPG}
	\end{axis}
\end{tikzpicture}
\begin{tikzpicture}
	\begin{axis}[width=0.38\textwidth,height=0.4\textwidth,
		xmajorgrids=true, xminorgrids=true,ymajorgrids=true,yminorgrids=true,grid style=dashed,xtick={1,3,5},ytick={20,21,22},
		xmin=0,xmax=5,ymin=19,ymax=23,xlabel=$t$ (s), ylabel=$v$ (m/s),]			
		\addplot[smooth,color=Fuchsia,very thick] table [x=T,y=V]{Data/CM_DOOT_REF.dat};
		\addplot[smooth,color=Dandelion,very thick] table [x=T,y=V]{Data/CM_DOOT_IPG.dat};
	\end{axis}
\end{tikzpicture}
\subcaption{Double overtake}\label{fig:ipg3}\vspace{5pt}
\end{subfigure}			
\begin{subfigure}[b]{\linewidth}\centering
\begin{tikzpicture}
	\begin{axis}[width=0.62\textwidth,height=0.4\textwidth,ytick={-5,0,5},legend columns=2,
		legend style={draw=none,fill=none,at={(0.45,0.9)},anchor=north},
		xmin=220,xmax=350,ymin=-6,ymax=6,xlabel=$x\e$ (m), ylabel=$y\e$ (m),]	
		\draw[gray,dashed] (200,0)--(400,0);		
		\draw[gray] (200,5)--(400,5);		
		\draw[gray] (200,-5)--(400,-5);		
		\addplot[smooth,color=Fuchsia,very thick] table [x=X,y=Y]{Data/CM_LACH_REF.dat};
		\addplot[smooth,color=Dandelion,very thick] table [x=X,y=Y]{Data/CM_LACH_IPG.dat};
		\legend{Planner, IPG}
	\end{axis}
\end{tikzpicture}
\begin{tikzpicture}
	\begin{axis}[width=0.38\textwidth,height=0.4\textwidth,
		xmajorgrids=true, xminorgrids=true,ymajorgrids=true,yminorgrids=true,grid style=dashed,xtick={1,3,5},ytick={20,21,22},
		xmin=0,xmax=5,ymin=19,ymax=23,xlabel=$t$ (s), ylabel=$v$ (m/s),]
		\addplot[smooth,color=Fuchsia,very thick] table [x=T,y=V]{Data/CM_LACH_REF.dat};
		\addplot[smooth,color=Dandelion,very thick] table [x=T,y=V]{Data/CM_LACH_IPG.dat};
	\end{axis}
\end{tikzpicture}
\subcaption{Lane change}\label{fig:ipg4}\vspace{5pt}
\end{subfigure}			
\caption{Simulation results for attainability assessment of the P--SMPC planner (Section~\ref{sec:attain}).}\label{fig:ipgsim}
\end{center}
\end{figure}
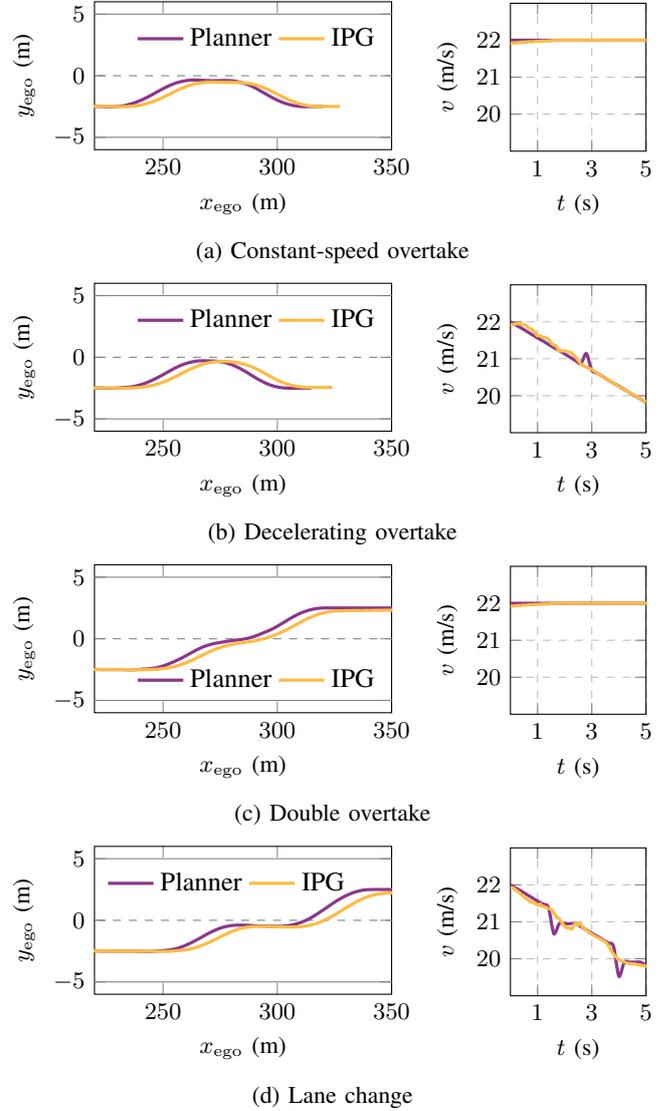

\subsection{Comparison with IPG Motion Planner}\label{sec:comparative}

As the final step, we showcase the proactivity and efficiency of the P--SMPC planner by comparing its behavior against the built-in collision avoidance module in IPG CarMaker simulation environment. The test scenario is similar to the complex scenario in Fig.~\ref{fig:scen4} where one static and four slow-moving obstacles are present on the road. This time, we decrease the obstacle velocities even further down to 2-7 m/s. Moreover, we simulate a sudden braking by the last obstacle on the road until it stops in a dangerous way. 

For a fair comparison, we set a ``normal" but ``risk-taking" driver behavior in IPG by selecting a standard driver and the maximum overtaking rate, which means that the driver always favors evading the obstacles rather than braking. This case shows how an overly-conservative planning strategy can lead to higher risk and propagating the hazard to other road users.

In the proactive collision avoidance case, we first ran the simulation in \textsc{Matlab} and used the same TestRun in CarMaker. We arrange the maneuver in IPG such that the IPG driver merely tracks the speed profile and the steering wheel angle generated by the P--SMPC planner in \textsc{Matlab}. Notably, we intentionally excluded considerations of other traffic participants in this scenario to prevent any interference with the operation of the IPG motion planner.

The video of the comparison simulation is accessible online\footnote{\url{https://youtu.be/UacmQDjQ2vI}}. Figure~\ref{fig:ipgvel} compares the velocity profiles for the overly-conservative IPG motion planner and the P--SMPC planner. While the P--SMPC planner manages to keep the velocity close to the cruising speed, the IPG planner dangerously brakes in multiple occasions. This issue becomes more critical when the IPG planner decides for a full stop behind the last obstacle on the right lane as shown in Fig.~\ref{fig:cshots3}: on the other hand, the P--SMPC planner manages to safely guide the ego vehicle outside of the risky zone between two slow-moving vehicles by taking a proactive strategy to overtake the stopping vehicle as well as by keeping a safe distance from the other slow-moving obstacle on the left lane in Fig.~\ref{fig:pshots3}. 
\begin{figure}[hbtp]
\begin{center}
\begin{subfigure}[b]{\linewidth}\centering
\begin{tikzpicture}
	\begin{axis}[width=\textwidth,height=0.6\textwidth, 
		xmajorgrids=true, xminorgrids=true,ymajorgrids=true,yminorgrids=true,grid style=dashed,ytick={0,5,10,15,20,25},legend columns=2,legend style={fill=white,draw=none},
		xmin=225,xmax=1000,ymin=0,ymax=27,xlabel=$x\e$ (m), ylabel=$v$ (m/s),]			
		\addplot[smooth,color=Fuchsia,very thick] table [x=X,y=V]{Data/CM_Complex_REF.dat};
		\addplot[smooth,color=Dandelion,very thick] table [x=X,y=V]{Data/CM_Complex_IPG.dat};
		\legend{P--SMPC, IPG}
	\end{axis}
\end{tikzpicture}
\subcaption{Velocity profiles of the ego vehicle along the road}\label{fig:ipgvel}\vspace{5pt}
\end{subfigure}\vspace{5pt}		
\begin{subfigure}[b]{\linewidth}\centering
\begin{tikzpicture}
	\begin{axis}[width=\textwidth,height=0.6\textwidth,
		xmajorgrids=true, xminorgrids=true,ymajorgrids=true,yminorgrids=true,grid style=dashed,legend columns=1,legend style={fill=white,draw=none},legend pos=north west,
		xmin=0,xmax=40,ymin=0,ymax=0.005,xlabel=$t$ (s), ylabel=Probability/Risk,]			
		\addplot[color=Maroon,very thick] table [x=T,y=J]{Data/CM_Complex_REF.dat};
		\addplot[color=NavyBlue,very thick] table [x=T,y=C]{Data/CM_Complex_REF.dat};
		\legend{$P$,$\max\{[\mathbb{P}]\mmps\}$}
	\end{axis}
\end{tikzpicture}
\subcaption{Risk and maximum $[\mathbb{P}]\mmps$ values for P--SMPC}\label{fig:ipgjc}\vspace{5pt}
\end{subfigure}
\caption{Plots of comparative test between overly-conservative and proactive collision avoidance (Section~\ref{sec:comparative}).}\label{fig:ipgscene}
\end{center}
\end{figure}
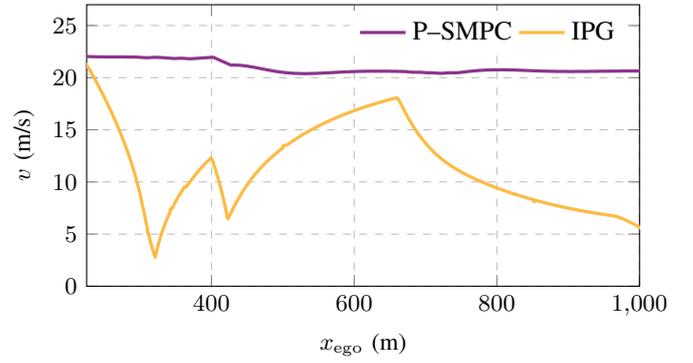
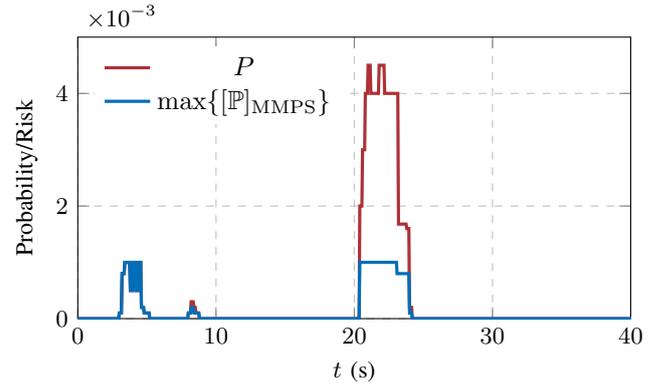

\begin{figure}[hbtp]
\begin{center}
\begin{subfigure}[b]{\linewidth}\centering
	\includegraphics[width=0.9\textwidth]{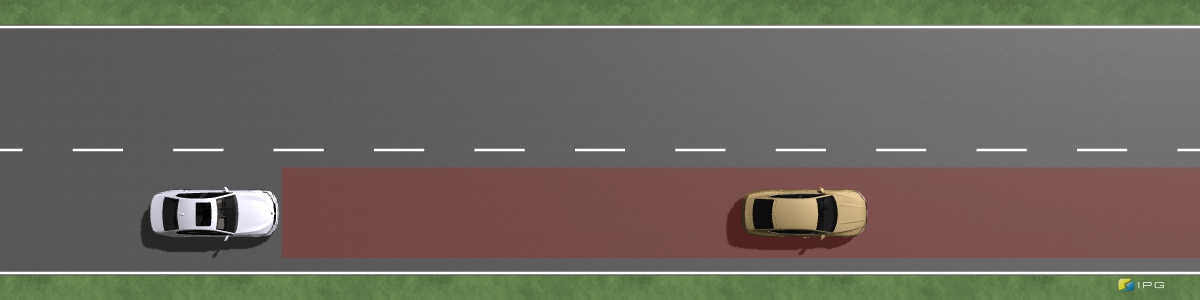}
	\includegraphics[width=0.9\textwidth]{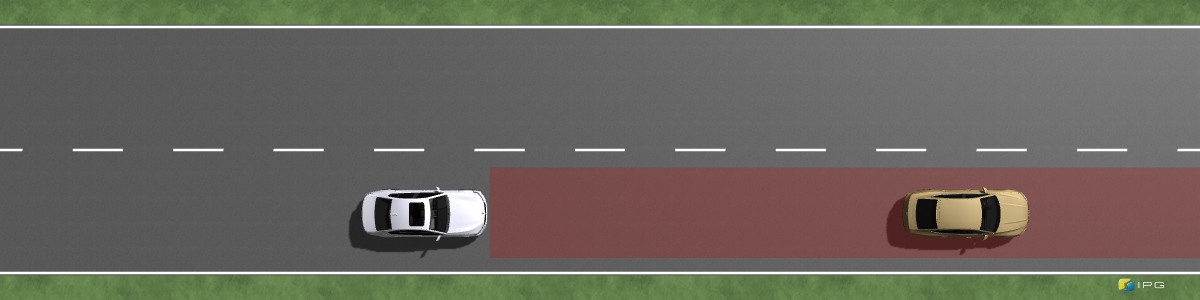}
	\includegraphics[width=0.9\textwidth]{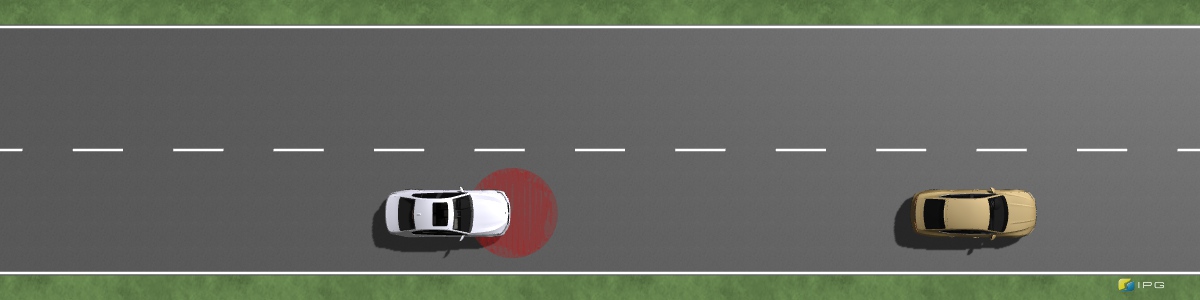}
	\subcaption{Overly-conservative collision avoidance: the ego vehicle slows down to keep distance until a full stop behind the obstacle.}\label{fig:cshots3}\vspace{5pt}
\end{subfigure}\vspace{5pt}				
\begin{subfigure}[b]{\linewidth}\centering
	\includegraphics[width=0.9\textwidth]{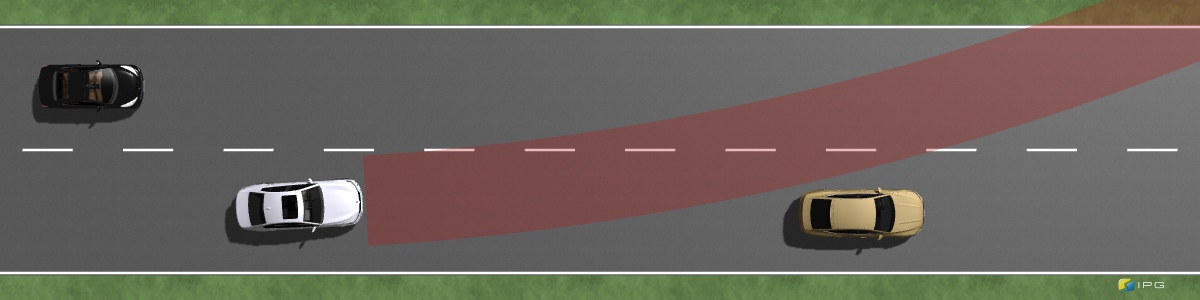}
	\includegraphics[width=0.9\textwidth]{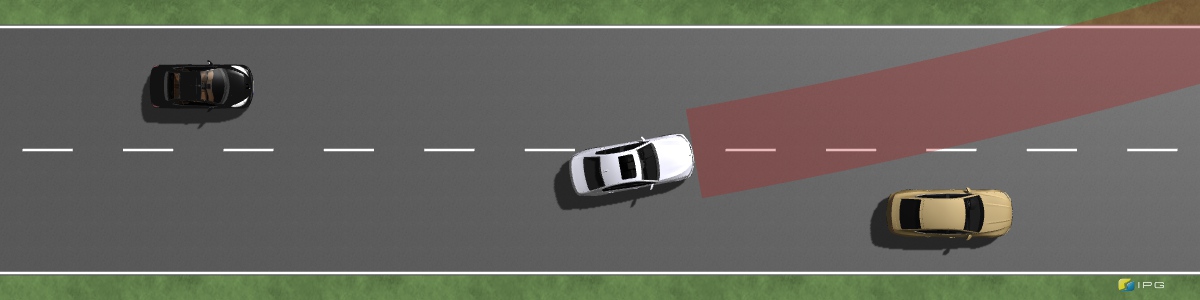}
	\includegraphics[width=0.9\textwidth]{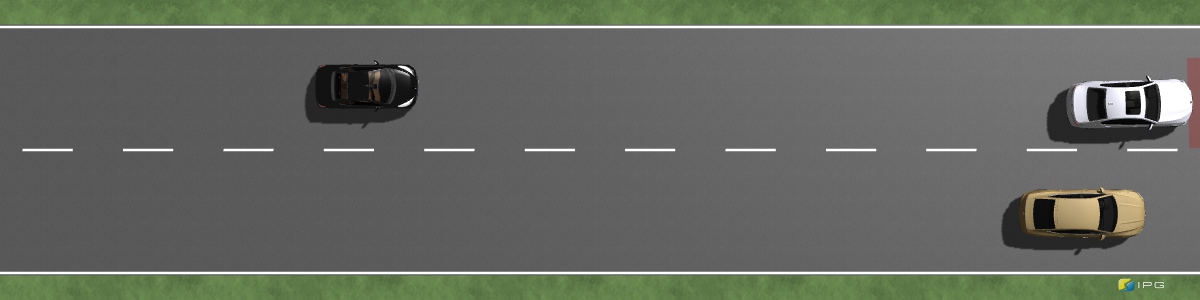}
	\subcaption{Proactive collision avoidance by P--SMPC: the ego vehicle manages to get out of the risky zone before its front vehicle stops.}\label{fig:pshots3}\vspace{5pt}
\end{subfigure}			
\caption{Snapshots of overly-conservative (upper) and proactive (lower) collision avoidance planning strategies (Section~\ref{sec:comparative}).}\label{fig:snapshots}
\end{center}
\end{figure}

\subsection{Performance Analysis and Discussion}

In the previous sections, we showed the proactivity of our proposed P--SMPC motion planner by comparing its performance against the state-of-the-art SMPC formulation (R--SMPC) and the built-in motion planner in a high-fidelity modeling and simulation platform. To gain a more clear view of the planning performance of P--SMPC, we have collected the data from all the aforementioned simulations and plotted the time evolution of chance constraints and the risk function values and the density histogram for computation time in Fig.~\ref{fig:costs}. Since the simulations have various lengths in terms of time, we have scaled their data to a risky zone and a safe zone in Figures~\ref{fig:costc} and~\ref{fig:costj} to allow for a meaningful comparison. The risky zone represents the section of the simulations where the ego vehicle observes sudden appearance of the obstacles and ends when it does not detect any obstacles ahead on the road.

Figure~\ref{fig:costc} shows the statistical information of $[\mathbb{P}]\mmps$ values in the Monte-Carlo simulation results for R--SMPC and P--SMPC planners. The maximum values for both planners is 0.001 (0.1\%) as shown in gray. Both planners show a reduction of the maximum $[\mathbb{P}]\mmps$ value by getting out of the risky zone. However, the mean for $[\mathbb{P}]\mmps$ values for P--SMPC are significantly lower than the mean values for R--SMPC, which shows the effectiveness of minimizing a risk function based on over-approximation of the $\mathbb{P}$ within the SMPC formulation. The peak in the mean value for P--SMPC corresponds to the riskiest time steps during the simulation, which occur where the vehicle is closest to the obstacle, e.g.\ during an overtaking maneuver. Further, the InterQuartile Range (IQR) distance for the planners is shown by the width of a shaded area around the mean values, using their corresponding colors.

The risk function values for P--SMPC planner are plotted in Fig.~\ref{fig:costj}. Since the risk function is an over-approximation of $\mathbb{P}$, its value are higher than $[\mathbb{P}]\mmps$. Nevertheless, the P--SMPC planner manages to keep the risk function below 0.0045 (0.45\%) at all times in Fig.~\ref{fig:ipgjc} due to its predictive proactive collision avoidance. In addition, while convergence to a global optimum cannot be guaranteed for an NLP, an MILP solver can reach its global optimum when it is given sufficient time. As a result, the MILP formulation of the (originally nonlinear) SMPC planning optimization problem improves the computational efficiency by a speed-up in computations and a better coverage of the decision space.

Lastly, the density histogram for computation time per planning step is shown in Fig.~\ref{fig:costtime}. Compared to the planner sampling time of $0.2$s, the MILP solver could find the global optimum 96\% of the times within $0.15$s (75\% of the time step) on our PC and only 4\% of the times required more than $0.2$s to find the global optimum. This shows the computational efficiency of the P--SMPC planning formulation, which can be further improved by imposing a time limit for the solver (and trading the global optimality) or running the simulations on a faster machine. Note that this level of computational efficiency is achieved for the assumed model and approximation accuracy adopted in this paper. For a more comprehensive study of control performance vs.~computational speed trade-off in hybridization of nonlinear MPC using MMPS formalism, the reader is referred to our previous study~\cite{HMPCPartI,HMPCPartII}.
\begin{figure}[htb]
\begin{center}
\begin{subfigure}[b]{0.5\textwidth}\centering
\begin{tikzpicture}
	\begin{axis}[small,ymin=0,height=0.6\textwidth,width=1.02\textwidth,
		ylabel={$[\mathbb{P}]\mmps$},xlabel=Simulation time/step,xmin=0,xmax=5,
		xtick={0,2.7,5},x tick label as interval,xticklabels={Risky zone, Safe zone},
		legend style={fill=none,draw=none},xmajorgrids=true,]
		\path[draw,color=gray,ultra thin,name path=axis] (0,0)--(5,0); 
		\addplot[smooth,color=gray,very thick,tension=0.4,name path=rmax] table [x=T,y=RC_MAX]{Data/Costs.dat};
		\addplot[smooth,color=Maroon,thin,tension=0.4,name path=psigp,forget plot] table [x=T,y=P_SIGP]{Data/Costs.dat};
		\addplot[smooth,color=Maroon,thin,tension=0.4,name path=psigm,forget plot] table [x=T,y=P_SIGM]{Data/Costs.dat};
		\addplot[smooth,color=NavyBlue,thin,tension=0.4,name path=rsigp,forget plot] table [x=T,y=R_SIGP]{Data/Costs.dat};
		\addplot[smooth,color=NavyBlue,thin,tension=0.4,name path=rsigm,forget plot] table [x=T,y=R_SIGM]{Data/Costs.dat};
		\addplot[fill=gray,fill opacity=0.2,forget plot] fill between [of=axis and rmax];
		\addplot[fill=NavyBlue,fill opacity=0.2] fill between [of=rsigp and rsigm];
		\addplot[fill=Maroon,fill opacity=0.2] fill between [of=psigp and psigm];
		\addplot[smooth,color=NavyBlue,very thick,tension=0.4] table [x=T,y=RC_MEAN]{Data/Costs.dat};
		\addplot[smooth,color=Maroon,very thick,tension=0.4] table [x=T,y=J_MEAN]{Data/Costs.dat};
		\legend{Maximum value,R--SMPC IQR,P--SMPC IQR,R--SMPC Mean,P--SMPC Mean};
	\end{axis}
\end{tikzpicture}
\subcaption{Evolution of $[\mathbb{P}]\mmps$ values}\label{fig:costc}\vspace{5pt}
\end{subfigure}			
\begin{subfigure}[b]{0.5\textwidth}\centering
\begin{tikzpicture}
	\begin{axis}[small,ymin=0,height=0.6\textwidth,width=\textwidth,
		ylabel=Risk function,xlabel=Simulation time/step,xmin=0,xmax=5,
		xtick={0,2.7,5},x tick label as interval,xticklabels={Risky zone, Safe zone},legend style={fill=none,draw=none},
		xmajorgrids=true]
		\path[draw,color=gray,ultra thin,name path=axis] (0,0)--(5,0); 
		\addplot[smooth,color=gray,very thick,tension=0.4,name path=rmax] table [x=T,y=J_MAX]{Data/Costs.dat};
		\addplot[smooth,color=Mulberry,thin,tension=0.4,name path=psigp,forget plot] table [x=T,y=J_SIGP]{Data/Costs.dat};
		\addplot[smooth,color=Mulberry,thin,tension=0.4,name path=psigm,forget plot] table [x=T,y=J_SIGM]{Data/Costs.dat};
		\addplot[fill=Mulberry,fill opacity=0.2] fill between [of=psigp and psigm];
		\addplot[smooth,color=Mulberry,very thick,tension=0.4] table [x=T,y=J_MEAN]{Data/Costs.dat};
		\legend{Maximum value,IQR,Mean value};
	\end{axis}
\end{tikzpicture}
\subcaption{Evolution of the risk function values (P--SMPC)}\label{fig:costj}\vspace{5pt}
\end{subfigure}					
\begin{subfigure}[b]{0.5\textwidth}\centering
\begin{tikzpicture}
	\begin{semilogxaxis}[small,ymin=0,height=0.6\textwidth,width=\textwidth,
		ylabel=Frequency density,xlabel=Computation time per step (s),xtick={0.1,0.2,0.5,1,2,3.5},log ticks with fixed point,legend style={fill=none,draw=none},]
		\draw[black,dashed] (0.2,0)--(0.2,12) node[midway,below,sloped] {\footnotesize Sampling time};	
		\addplot+[hist={density,bins=31},fill=BurntOrange!50,draw=BurntOrange!75!black,mark=no] table [y index=0] {Data/Times_NL.dat};
		\addplot+[hist=density,fill=OliveGreen!50,draw=OliveGreen!75!black,mark=no,fill opacity=0.7] table [y index=0] {Data/Times_MMPS.dat};
		\legend{P--SMPC,N--SMPC};
	\end{semilogxaxis}
\end{tikzpicture}
\subcaption{Density histogram for computation times. The N--SMPC computation times for steps that the NLP was infeasible are not considered and the data only account for the duration of sampling times where the planner converged to a solution.}\label{fig:costtime}\vspace{5pt}
\end{subfigure}					
\caption{Performance analysis of the P--SMPC planner in terms of safety and computation time. The data in these plots represent the density histograms of their corresponding variables considering all the performed simulations in this study.}\label{fig:costs}
\end{center}
\end{figure}
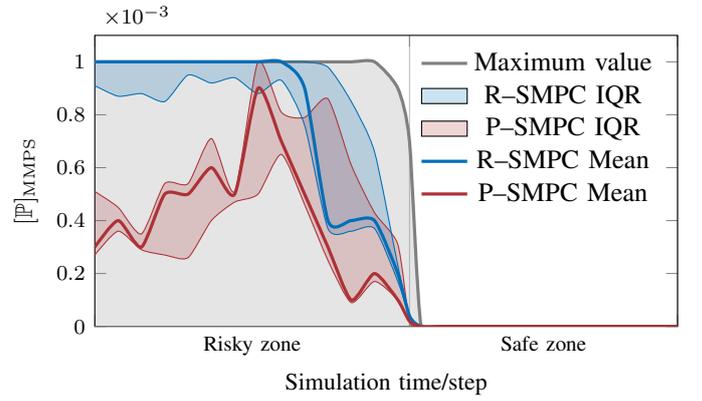
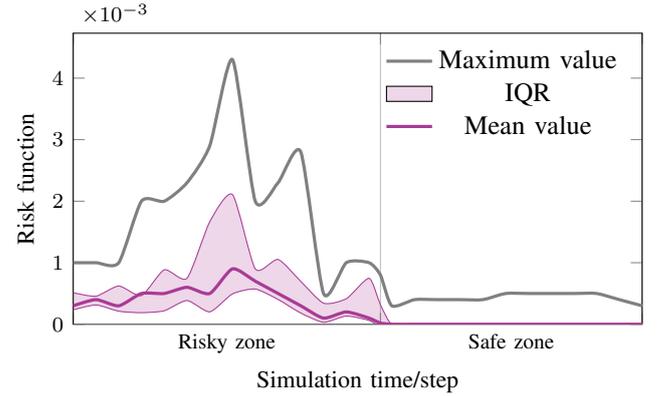
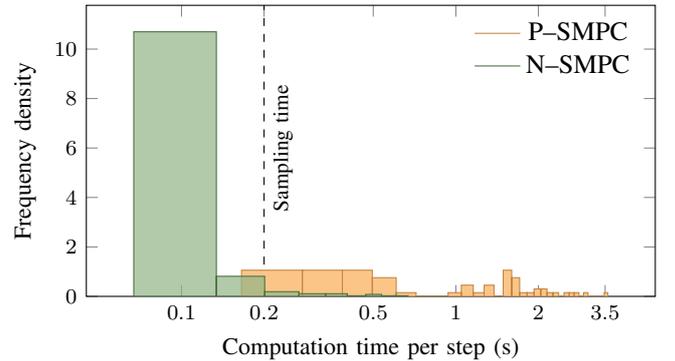


\section{Conclusions}\label{sec:conc}

This paper has presented a novel SMPC motion planner for emergency collision avoidance during hazardous highway scenarios. The proposed planner proactively avoids collision by static and dynamic obstacles on a highway by avoiding conservatism and swift response to sudden appearance of road users with uncertain behavior, thus improving the safety of the ego vehicle. 

The novelties of our proposed approach can be summarized as follows: first, the proactive SMPC planner uses a tractable formulation of chance constraints for safe collision avoidance, while minimizing a risk function formulated as an over-estimation of the probabilities while facilitating the incorporation of a dynamic model for the ego vehicle as well as exploiting the tire-force potential close to the vehicle handling limits. Secondly, hybrid approximations of the nonlinearities in the system dynamics by MMPS formalism are used to allow for an MILP formulation of the SMPC problem and facilitate real-time implementation and convergence to the global optimum. Safety, proactivity, and computational efficiency of our proposed planned were shown via various simulations of emergency scenarios and compared against the state-of-the-art SMPC formulation and a high-fidelity vehicle modeling and simulation environment. 

For future work, we aim at improving the model for dynamic obstacles on the road and extending the uncertainty regarding the intention of the other road users. While the model employed in this paper for the obstacles helped obtain an efficient computational accuracy-speed trade-off, more comprehensive models of obstacle behavior are influential for implementation of levels 4 and 5 of automated driving. Further, we aim at integrated planning and control design for emergency scenarios for improved accuracy and computational efficiency, in addition to investigating an efficient control structure to integrate our proposed SMPC planner with hybrid vehicle control and a friction estimator to account for the uncertainties of the environment as well. Moreover, in-depth calibration of probability bounds, investigation of suboptimality bounds, feasibility analysis of the SMPC problem for different probability formulations, and proof of recursive feasibility will be important topics for our future research, as well as designing a back-up mode in cases where the feasibility of the planning optimization problem cannot be guaranteed.


\section*{Acknowledgments}

This research is funded by the Dutch Science Foundation NWO-TTW within the EVOLVE project (no.\ 18484).


\bibliographystyle{IEEEtran}
\bibliography{Citations}

\vspace*{-1.5cm}
\begin{IEEEbiography}[{\includegraphics[width=1in,height=1.25in,clip,keepaspectratio]{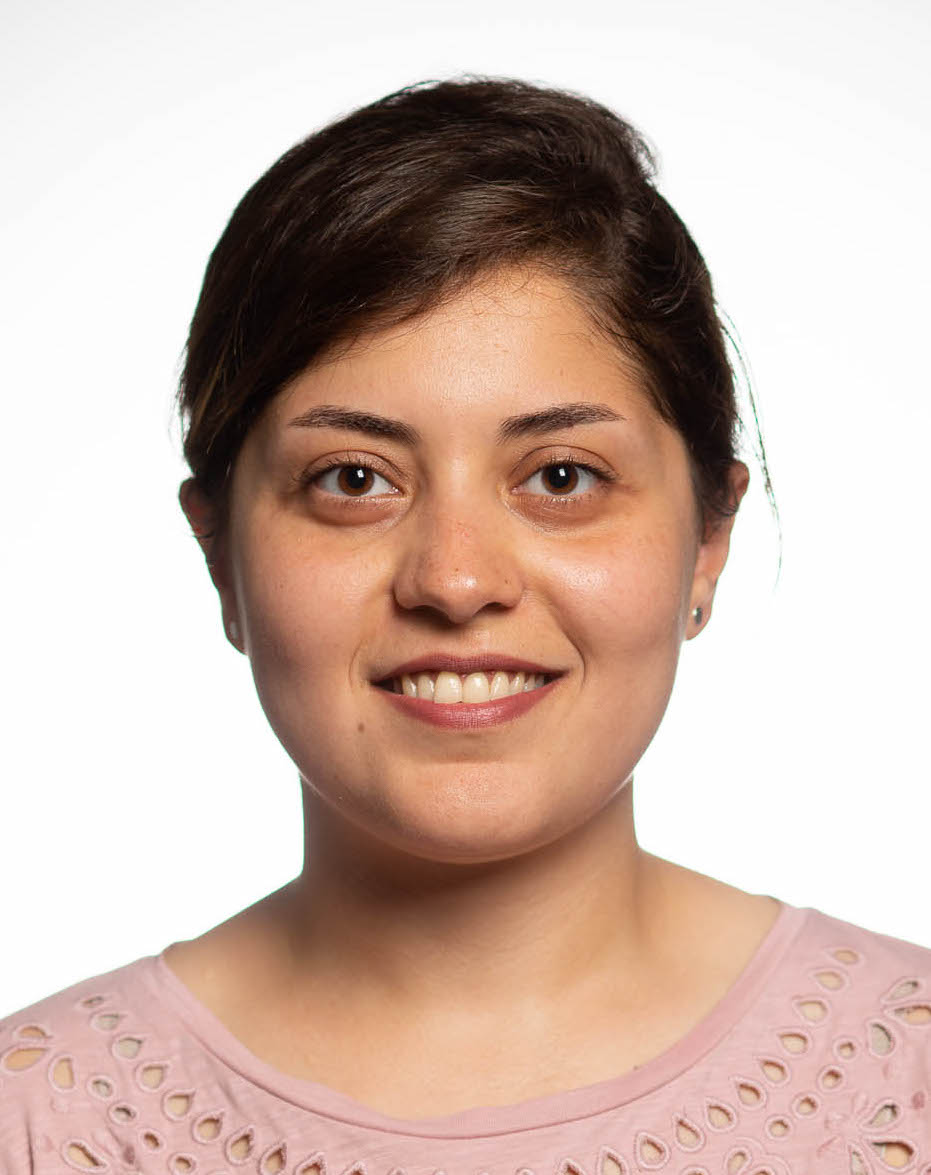}}]{Leila Gharavi} 
	is a PhD candidate at Delft Center for Systems and Control, Delft University of Technology, The Netherlands. She received her BSc and MSc degrees in mechanical engineering from Amirkabir University of Technology (Tehran Polytechnic) in Iran and has research experience in automatic manufacturing and production, vibration analysis and control of nonlinear dynamics, and soft rehabilitation robotics. 
	
	Currently, her research focuses on nonlinear and hybrid systems, optimization, and model-predictive control, with applications to adaptive and proactive control of automated  vehicles in hazardous scenarios.
\end{IEEEbiography}
\vspace*{-1.5cm}
\begin{IEEEbiography}[{\includegraphics[width=1in,height=1.25in,clip,keepaspectratio]{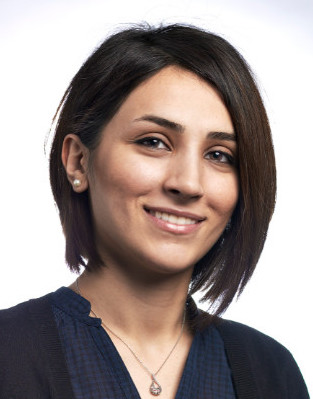}}]{Azita Dabiri} 
	received the Ph.D. degree from the Automatic Control Group, Chalmers University of Technology, in 2016. She was a Post-Doctoral Researcher with the Department of Transport and Planning, TU Delft, from 2017 to 2019. In 2019, she received an ERCIM Fellowship and also a Marie Curie Individual Fellowship, which allowed her to perform research at the Norwegian University of Technology (NTNU), as a Post-Doctoral Researcher, from 2019 to 2020, before joining the Delft Center for Systems and Control, TU Delft, as an Assistant Professor. Her research interests are in the areas of integration of model-based and learning-based control and its applications in transportation networks.
\end{IEEEbiography}
\vspace*{-1.5cm}
\begin{IEEEbiography}[{\includegraphics[width=1in,height=1.25in,clip,keepaspectratio]{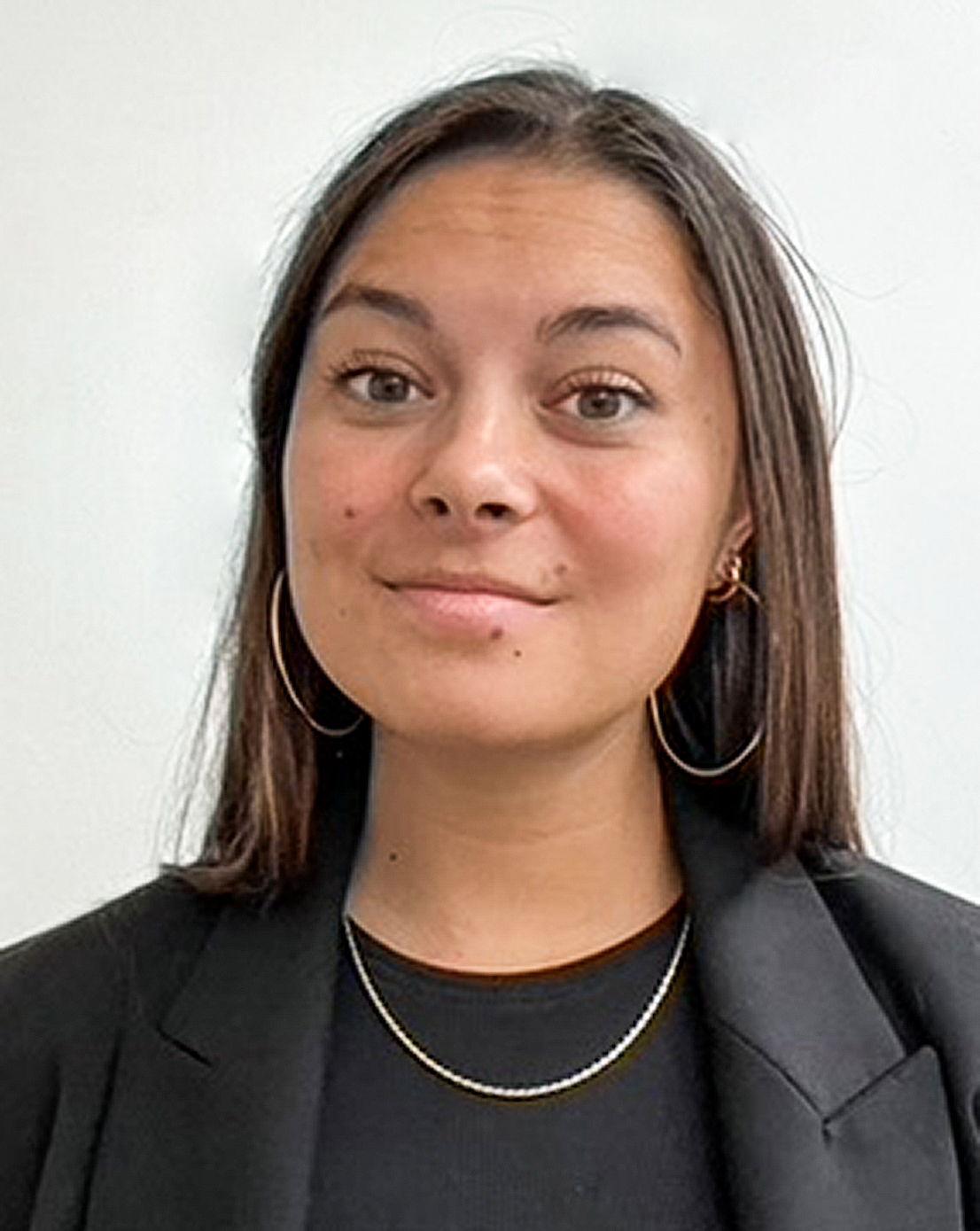}}]{Jelske Verkuijlen} 
	received the B.Sc. degree in Mechanical Engineering from Delft University of Technology in 2020. During her undergraduate years, she served as a drivetrain engineer at Formula Student Team Delft. Currently, she is pursuing her M.Sc. degree in Systems \& Control at Delft Center for Systems and Control, Delft University of Technology, where her research interests primarily revolve around stochastic predictive control and its applications in automated driving.
\end{IEEEbiography}
\vspace*{-1.5cm}
\begin{IEEEbiography}[{\includegraphics[width=1in,height=1.25in,clip,keepaspectratio]{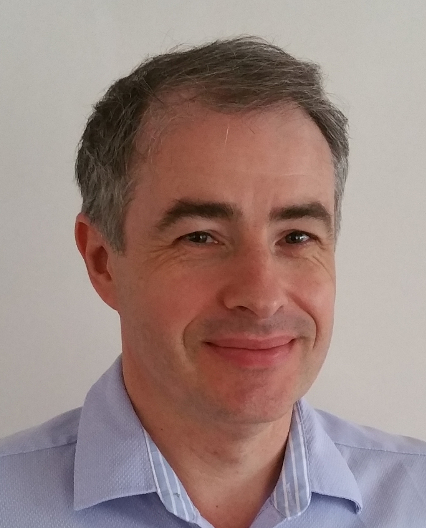}}]{Bart De Schutter}(Fellow, IEEE) 
	received the PhD degree (\emph{summa cum laude}) in applied sciences from KU Leuven, Belgium, in 1996. He is currently a Full Professor and Head of Department at the Delft Center for Systems and Control, Delft
	University of Technology, The Netherlands. His research interests include multi-level
	and multi-agent control, model predictive control, learning-based control, and control
	of hybrid systems, with applications in intelligent transportation systems and smart energy systems. 
	
	Prof.\ De Schutter is a Senior Editor of the IEEE Transactions on Intelligent Transportation Systems and an Associate Editor of the IEEE Transactions
	on Automatic Control.
\end{IEEEbiography}
\vspace*{-1.5cm}
\begin{IEEEbiography}[{\includegraphics[width=1in,height=1.25in,clip,keepaspectratio]{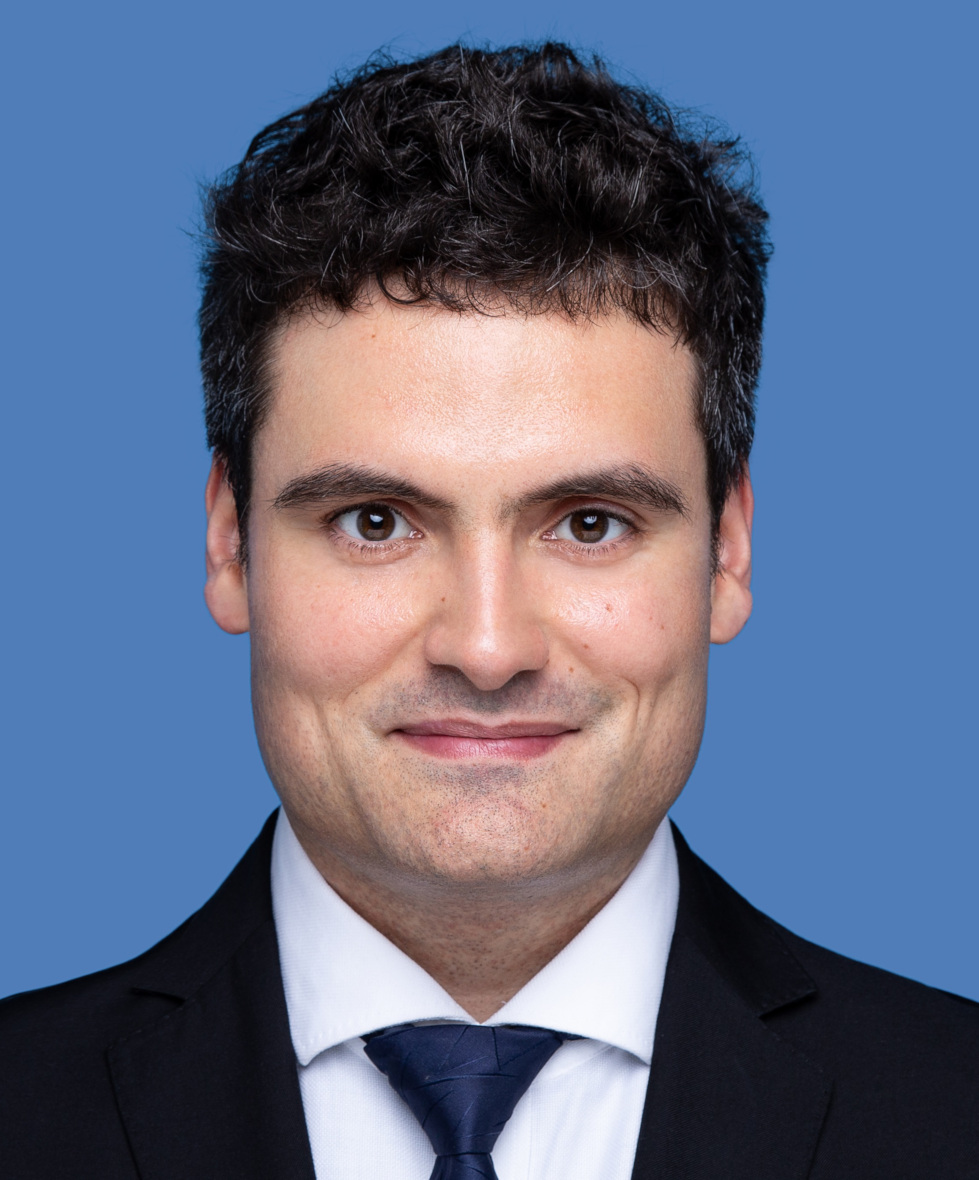}}]{Simone Baldi}
	(Senior Member, IEEE) received the B.Sc. in electrical engineering, and the M.Sc. and Ph.D. in automatic control engineering from University of Florence, Italy, in 2005, 2007, and 2011, respectively. Since 2019, he is a professor with Southeast University, China, previously being assistant professor with Delft Center for Systems and Control, Delft University of Technology, The Netherlands. His research interests include adaptive and learning systems with applications in intelligent vehicles and smart energy. He was awarded outstanding reviewer of Applied Energy in 2016, Automatica in 2017, AIAA Journal of Guidance, Control, and Dynamics in 2021. He is a Subject Editor of International Journal of Adaptive Control and Signal Processing, a Technical Editor of IEEE/ASME Transactions on Mechatronics, and an Associate Editor for IEEE Control Systems Letters and Journal of the Franklin Institute.
\end{IEEEbiography}

\end{document}